\documentclass[aps,prd,showpacs,amssymb,floatfix,nofootinbib,superscriptaddress]{revtex4-1}
\usepackage{graphicx,amsmath,subfigure}

\newcommand{\eeq}{\end{equation}}
\newcommand{\bea}{\begin{eqnarray}}

\def\ltsima{$\; \buildrel < \over \sim \;$}
\def\simlt{\lower.5ex\hbox{\ltsima}}
\def\gtsima{$\; \buildrel > \over \sim \;$}
\def\simgt{\lower.5ex\hbox{\gtsima}}

\newcommand{\hGR}{\mathbf{h}_\mathrm{GR}}
\newcommand{\hAP}{\mathbf{h}_\mathrm{AP}}
\newcommand{\tbf}{\theta}
\newcommand{\ttr}{\hat \theta}

\def\lesssim{\mathrel{\hbox{\rlap{\hbox{\lower4pt\hbox{$\sim$}}}\hbox{$<$}}}}
\def\gtrsim{\mathrel{\hbox{\rlap{\hbox{\lower4pt\hbox{$\sim$}}}\hbox{$>$}}}}
\def\alt{\mathrel{\hbox{\rlap{\hbox{\lower4pt\hbox{$\sim$}}}\hbox{$<$}}}}
\def\agt{\mathrel{\hbox{\rlap{\hbox{\lower4pt\hbox{$\sim$}}}\hbox{$>$}}}}
\def\PRD{{\it Phys. Rev.} D~}
\def\PRL{{\it Phys.Rev.} Lett~}

\def\gta{\ifmmode {\mathbin{\lower 3pt\hbox   
    {$\,\rlap{\raise 5pt\hbox{$\char'076$}}\mathchar"7218\,$}}}
    \else {${\mathbin{\lower 3pt\hbox
    {$\rlap{\raise 5pt\hbox{$\char'076$}}\mathchar"7218\,$}}}
    $}\fi}
\def\lta{\ifmmode {\,\mathbin{\lower 3pt\hbox   
    {$\,\rlap{\raise 5pt\hbox{$\char'074$}}\mathchar"7218\,$}}}
    \else {${\mathbin{\lower 3pt\hbox
    {$\rlap{\raise 5pt\hbox{$\char'074$}}\mathchar"7218\,$}}}
    $}\fi}

\begin{document}
\title{Importance of including small body spin effects in the modelling of extreme and intermediate mass-ratio inspirals}

\author{E.A. Huerta}
\affiliation{Institute of Astronomy, Madingley Road, CB3 0HA Cambridge, UK}
\affiliation{Department of Physics, Syracuse University, Syracuse, NY 13244, USA.}
\author{Jonathan R.  Gair}
\affiliation{Institute of Astronomy, Madingley Road, CB3 0HA Cambridge, UK}

\email{eah41@ast.cam.ac.uk}
\email{jgair@ast.cam.ac.uk}


\date{\today}

\begin{abstract}        
In this paper we explore the ability of future low--frequency gravitational wave detectors to measure the spin of stellar mass and intermediate mass black holes (IMBHs) that inspiral into spinning supermassive black holes (SMBHs). We describe a kludge waveform model based on the equations of motion derived by Saijo et al. \cite{maeda} for spinning BH binaries, augmented with spin--orbit and spin--spin couplings taken from perturbative and post--Newtonian (PN) calculations, and the associated conservative self--force corrections, derived by comparison to PN results. We model the inspiral phase using accurate fluxes which include perturbative corrections for the spin of the inspiralling body, spin--spin couplings and higher--order fits to solutions of the Teukolsky equation. We present results of Monte Carlo simulations of parameter estimation errors, computed using the Fisher Matrix formalism, and also of the model errors that arise when we omit conservative corrections from the waveform template. We present results for the inspirals of spinning BHs with masses \(\mu =10M_{\odot},\,10^2M_{\odot},\,10^3M_{\odot},\,5\times 10^3M_{\odot}\), into SMBHs of specific spin parameter \(q=0.9\), and mass \(M=  10^6M_{\odot}\). The analysis shows that for intermediate-mass-ratio inspirals, e.g., a source \(5\times10^3M_{\odot}+10^6M_{\odot}\) observed with a signal-to-noise ratio (SNR) of 1000,  observations made with LISA will be able to determine the inspiralling BH mass, the central SMBH mass, the SMBH spin magnitude,  and the magnitude of the spin of the inspiralling BH to within fractional errors of \(\sim 10^{-3},\,10^{-3},\,10^{-4}\), \( 10\%\), respectively. We also expect to determine the location of the source in the sky and the SMBH spin orientation to within \(\sim 10^{-4}\) steradians.  However, for extreme-mass-ratio inspirals, e.g., a \(10M_{\odot}+10^6M_{\odot}\) system observed with SNR of 30, LISA will not be able to determine the spin magnitude of the inspiralling object, although the measurement of the other waveform parameters is not significantly degraded by the presence of spin. We also show that the model errors which arise from ignoring conservative corrections become significant for mass-ratios above $\sim10^{-4}$, but including these corrections up to second PN order may be sufficient to reduce these systematic errors to an acceptable level.

\end{abstract}


\maketitle

\section{Introduction}    
Future low--frequency gravitational wave (GW) detectors, such as LISA~\cite{noise}, will provide information about the astrophysical properties of a great variety of sources, including merging SMBHs, stellar mass compact object (CO) binaries in our galaxy and the inspirals of stellar mass COs into SMBHs in the centres of galaxies. The scientific payoffs of such observations will shed light on the spin and mass distributions of SMBHs, the mass distributions of inspiralling stellar-mass objects in galactic centres and the cosmological rate of mergers, and unprecedented information on the properties and evolution of BHs in the Universe.  

In addition to providing this unique new information about astrophysical objects in the Universe, detectors like LISA will enable us to carry out tests of general relativity in the strong--field regime close to the horizon of massive BHs. These tests will allow us to determine whether the geometry of the BH is the Kerr metric of relativity, as we expect, or if it is some other kinds of massive body, such as a boson star, naked singularity, soliton star, etc. The extreme-mass-ratio inspirals (EMRIs) of stellar-mass compact objects into much more massive SMBHs in the centres of galaxies are ideally suited for such tests, as many hundreds of thousands of waveform cycles are generated while the smaller body is orbiting in the strong field regime close to the SMBH.

Ryan \cite{ryan} showed, for nearly-circular nearly-equatorial EMRIs, that the phase of the GWs observed will provide a map of the spacetime structure outside the central BH. Such a map might also be extracted for eccentric-inclined EMRIs \cite{nonkerr}, but with a few additional complications, since the eccentricity and inclination cannot be directly determined, but only inferred from the observed frequencies \cite{hair}. These maps can be compared to the Kerr metric in order to detect deviations from the general relativistic predictions. If deviations are observed, these could arise from a difference in the nature of the central object or a failure in the theory of relativity, but they might also have an astrophysical origin, e.g., the presence of gas in the system. However, most massive holes are in galaxies with non--active nuclei, and will be surrounded by only a tenuous disk with advection--dominated accretion flow. Narayan \cite{ramesh} has shown that in such environments accretion drag should be totally negligible, and hence the orbital evolution will be entirely driven by radiation reaction. Massive BHs in active galactic nuclei will be surrounded by thin accretion disks. In that case, accretion--disk drag will be significant, which could leave an imprint on the EMRI signal, but in general such astrophysical effects are qualitatively different from the effect of a deviation in the spacetime structure, e.g., they can drive increasing eccentricity~\cite{vac}.

If we are to detect small differences in the gravitational wave phasing, it is necessary to develop accurate waveform templates for the pure Kerr-inspiral model, against which to test the observations. Great progress has been made in the modelling of EMRI systems and the understanding gained from this work has driven the development of  algorithms for EMRI detection and data analysis \cite{mock34,ltf,mock1B3,GMH,ITFG,GWen}. The formalism to compute accurate signals is already well developed. Teukolsky \cite {saul} and Sasaki et al. \cite{sasamu} developed a mathematical framework to study first--order perturbations of Kerr BHs. However, this approach is computationally expensive and has left a need for the development of computationally inexpensive waveform templates that capture the main features of the true signals, and that can be used for source detection in data analysis algorithms. In this paper we will use one particular approach, the so--called ``numerical kludge'' waveform model.  This approach combines a prescription to capture the radiative dynamics of Teukolsky--based evolutions \cite{kludge}, using accurate expressions for the fluxes of energy and angular momentum to compute the inspiral trajectories with an approximate flat-spacetime model to construct the waveform from the trajectory~\cite{improved}. It has been shown that this approach gives waveforms whose overlap with fully accurate general relativistic waveforms is greater than \(95\%\) over a large portion of the parameter space \cite{kludge}. This approach has recently been improved for the case of circular-equatorial EMRIs, by including conservative self--force corrections at 2PN order \cite{cons}. These corrections are important because they affect the phasing of the waveform over time, and neglecting them may lead to several cycles of discrepancy in the waveform template over the course of  an inspiral. The self--force program has not yet obtained general relativistic conservative self--force corrections for Kerr inspirals, so in \cite{cons} these corrections were derived by comparison to PN results and requiring that the ``numerical kludge'' model reproduced PN results in the weak field.  Using this improved waveform model, we showed that the inclusion of such corrections in the model was not essential for source detection, but we may have to include them for accurate parameter determination for EMRIs. In this paper, we consider another effect, namely the influence of the spin of the inspiralling body on Kerr circular-equatorial inspirals and assess the importance of including this parameter in EMRI models. 

This analysis is important because it is expected that most astrophysical BHs will have significant spin and therefore the true gravitational waveforms will include small body spin effects. It has been argued that such effects may not play a significant role for signal detection and parameter estimation for EMRIs with mass ratios \(\eta=\mu/M \sim 10^{-5}\) \cite{cutler}. This is because the spin of the inspiralling object is suppressed by a factor of \(\eta\) in the equations of motion \cite{maeda}, and hence the inclusion of small body spin effects will modify the orbital evolution by at most a few radians over a year. Studies on the effect of the self--force on spinning objects that inspiral into Schwarzschild BHs have shown that the accurate determination of the local self--force may not be needed for a determination of the orbital evolution of EMRIs, and its omission will only introduce a small error in the determination of the spin rate of the companion \cite{burko}. However, the spin effects become more important as the mass ratio, $\eta$, is increased and can therefore be significant for intermediate mass ratio inspirals (IMRIs) with $\eta \sim 10^{-3}$ (e.g., the inspiral of a $\sim10^3M_{\odot}$ intermediate mass BH (IMBH) into a $\sim10^6M_{\odot}$ SMBH). Even if small body spin effects are only marginally important for detection, their inclusion in waveform templates may be important for parameter estimation. We also need to determine at what mass ratio small body spin effects become important for signal detection, i.e., when GW observations will be able to measure the spin of the inspiralling body. We address this question here by developing a ``numerical kludge'' waveform model that includes small body spin effects using the mathematical machinery developed by Saijo et al. \cite{maeda}, and by including conservative self--force corrections for spin--orbit and spin--spin couplings using the same method that was employed to perturbative conservative corrections in~\cite{cons}. We compute, for the first time,  the mass--ratio threshold \(\eta\)  above which low-frequency gravitational wave detectors like LISA will be able to accurately measure the spin of a BH inspiralling into a SMBH. We will show that future observations will not only provide an unprecedented census of the spin distribution of SMBHs, but could also yield an accurate census of the spin and mass distributions of intermediate mass BHs with mass \(\mu \gtrsim 10^3 M_{\odot}\). 

This paper is organized as follows. In Section~\ref{s2} we introduce the waveform model used to study the inspiral of spinning BHs into SMBHs. We construct kludge asymptotic observables to implement the radiative and conservative components of the self--force, including spin--orbit and spin--spin couplings, by comparison to PN results. Section~\ref{s3} presents the wave generation prescription used in our studies, along with the implementation of LISA's response function and Doppler phase modulations in our waveform model. In Section~\ref{s4} we review the basic elements of signal analysis, which we use in Section~\ref{s5} to assess the accuracy with which LISA observations might be able to determine the parameters of a representative sample of binary systems. In Section~\ref{s6} we describe the formalism developed by Cutler \& Vallisneri \cite{vallisneri} to estimate the theoretical or ``model'' errors that would arise from omitting conservative corrections in the waveform template. We summarize our conclusions in Section~\ref{s7}.

\section{Kludge waveform with small body spin effects}
\label{s2}

The numerical kludge waveform model developed by Babak et al. ~\cite{kludge} has been very successful as a model for the GW emission from EMRIs, as it is able to capture the main features of the inspiral waveform  in the strong field regime.  In this scheme, one assumes that the inspiralling object instantaneously follows a Kerr space-time geodesic, and slowly evolve the parameters of the geodesic using a prescription for the rate of change of energy, angular momentum and Carter constant that is based on PN expressions, augmented by fits to Teukolsky--based evolutions \cite{improved}. This results in a trajectory for the inspiralling body in Boyer--Lindquist coordinates. The gravitational waveform is then constructed by identifying these coordinates with spherical polar coordinates in a pseudo-flat space and using a weak-field emission formula. This model has already been used to explore the accuracy with which LISA will be able to measure the parameters of Kerr circular equatorial EMRIs \cite{cons}, and to estimate the importance of conservative and radiative self--force corrections on parameter estimation and detection. In the kludge models to date, the spin of the stellar mass CO was ignored. For the purpose of providing a better theoretical template, and to find out under which circumstances the spin of the CO can be measured, in this paper we will develop a new kludge model that incorporates this additional parameter.

To achieve this, we will augment the standard ``numerical kludge'' model using the equations of motion of a spinning particle in the equatorial plane ($\theta=\pi/2$) of a Kerr BH, as derived by Saijo et al. \cite{maeda}. We will include small body spin corrections of two different natures: i) first-order conservative corrections to amend the orbital phase evolution, and; ii) second order radiative corrections in the fluxes of energy and angular momentum to evolve the geodesic parameters of the inspiralling object. 

For a particle with spin angular momentum    ${\bf S}_1 = s \mu \hat{\bf z}$, aligned with the central Kerr BH spin (${\bf S}_2 = a M \hat{\bf z}$) and the orbital angular momentum $L_z$, the spin vectors remain constant and the equations of motion take a form similar to the Kerr geodesic equations, namely, \cite{maeda},

\begin{eqnarray}
\label{eq:papapetroueqns}
\label{eq:dtdtau}
\Sigma_s \Lambda_s \frac{dt}{d\tau} &=& a \left( 1 + \frac{3M s^2}{p \Sigma_s} \right) \! \left[ \tilde{J}_z - (a + s) \tilde{E} \right] \! + \! \frac{p^2+a^2}{\Delta}P_s, \\
\label{eq:dphidtau}
\Sigma_s \Lambda_s \frac{d\varphi}{d\tau} &=& \left( 1 + \frac{3M s^2}{p \Sigma_s} \right) \left[ \tilde{J}_z - (a + s) \tilde{E} \right] + \frac{a}{\Delta}P_s, \\
\label{eq:drdtau}
\Sigma_s \Lambda_s \frac{dp}{d\tau} &=& \pm \sqrt{R_s},
\end{eqnarray}

\noindent where,
\begin{eqnarray}
 \Sigma_s &=& p^2 \left( 1 - \frac{M s^2}{p^3} \right), \nonumber\\
\Lambda_s &=& 1- \frac{3 M s^2 p [\tilde{J}_z - (a+s)\tilde{E} ]^2}{\Sigma_s^3}, \nonumber\\
R_s &=&P_s^2 - \Delta \left\{ \frac{\Sigma_s^2}{p^2} + [ \tilde{J}_z - (a+s)\tilde{E} ]^2 \right\}, \nonumber\\
P_s &=& \left[ (p^2+a^2) + a s \left( 1 + \frac{M}{p} \right) \right] \tilde{E} - \left( a + s \frac{M}{p} \right) \tilde{J}_z, \nonumber\\
\Delta &=&p^2-2M p + a^2,
\end{eqnarray}

\noindent and $(t,p,\theta,\varphi)$ are Boyer-Lindquist coordinates, $\tau$ is the particle's proper time, and $\tilde{E}\equiv E/\mu$ and $\tilde{J}_z \equiv J_z/\mu$ are the conserved energy and total angular momentum, respectively (see Eqs.~(2.10) of \cite{maeda}). The radial motion of the spinning particle can be understood by re--writing the function $R_s$ in the form

\begin{equation}
R_s = B(p) [\tilde{E}-\tilde{E}_1(p,\tilde{J}_z)][\tilde{E}-\tilde{E}_2(p,\tilde{J}_z)],
\end{equation}
where the roots $\tilde{E}_{1,2}$ of $R_s=0$ are found by solving
\begin{equation}
\alpha \tilde{E}^2 - 2 \beta \tilde{E} + \gamma =0, \;\;\;\; \text{with}
\label{eneq}
\end{equation}
\begin{eqnarray}
\alpha &=& \left[ (p^2 + a^2) + a s \left( 1 + \frac{M}{p} \right) \right]^2 - \Delta (a + s)^2 , \\\nonumber
\beta &=& \! \left\{ \! \left( a + s \frac{M}{p} \right) \! \left[ \! (p^2 + a^2) \! + a s \left( \! 1 \! + \! \frac{M}{p} \right) \! \right] \! - \! \Delta (a \! + \! s) \! \right\} \! \tilde{J}_z, \\ \nonumber
\gamma &=& \left( a + s \frac{M}{p} \right)^2 \tilde{J}_z^2 - \Delta \left[ p^2 \left( 1 - \frac{M s^2}{p^3} \right)^2 + \tilde{J}_z^2 \right].
\end{eqnarray}

\noindent The effective potential is defined as the minimum allowed 
value of the particle energy at radius \(p\) and so Eq. \eqref{eneq} implies,
\begin{equation}
\label{eq:E1Veff}
V_{\rm eff} = \frac{\beta + \sqrt{\beta^2 - \alpha \gamma}}{\alpha}.
\end{equation}
\noindent Note that we have taken the positive square root because the motion of the spinning particle should be future directed.

We can now derive the expressions for the energy and angular momentum of a spinning particle, following a circular equatorial orbit by ensuring that the function \(R_s(p)\), and its radial gradient, \(R^{\,'}_s (p)\), vanish at the same point.  

The equations of motion derived by Saijo et al. \cite{maeda} are valid only to linear order in the spin of the small body, so we shall quote the expressions for the energy and angular momentum of the spinning particle at the same order, namely, 

\begin{eqnarray}
\frac{E}{\mu} &=& \frac{r^2 -2r \pm (q+\hat{s}/r)\sqrt{r+3q\hat{s}/r} - 5q \hat{s}/2r}{r\sqrt{r^2 -3r \pm (2q+3 \hat s/r)\sqrt{r+3q\hat{s}/r} - 6q\hat{s}/r}}, \nonumber \\   
\frac{L_{z}}{\mu M} &=& \frac{ \pm\sqrt{r+3q\hat{s}/r}\left(r^2 +q^2 +q\hat{s}(r+1)/r\right) -2rq+\hat s r(r-\frac{7}{2})}{r\sqrt{r^2 -3r \pm (2q+3 \hat{s}/r)\sqrt{r+3q\hat{s}/r} - 6q \hat{s}/r}},
\label{enlzsp}
\end{eqnarray}

\noindent where \(r=p/M\), \(\hat s= s/M =\eta \chi\), with \(\chi\) the dimensionless spin parameter of the inspiralling BH, \(\eta=\mu/M\) is the mass ratio as before, and \(q=a/M\), with  \(a\) the spin of the central BH.

Furthermore, for Schwarzschild BHs, \(q=0\), we find that at linear order in \(\hat s\), 

 \begin{eqnarray}
\frac{E}{\mu}\Bigg|_{q\rightarrow 0} &=& \frac{r-2}{\sqrt{r(r-3)}} -  \frac{\eta\chi}{2r (r-3)^{3/2}} + O(\hat s^2),
\label{ensp}\\
\frac{L}{\mu M}\Bigg|_{q\rightarrow 0} &=& \frac{r}{\sqrt{r - 3 }} + \frac{\eta\chi}{2} \frac{(r-2)(2r-9)}{\sqrt{r} (r-3)^{3/2}} + O(\hat s^2).
\label{amsp}
\end{eqnarray}

\noindent Eq. \eqref{ensp} corrects a typo in Eq. (B18) of \cite{favata}. 

To obtain the orbital evolution of the CO, we need to calculate the evolution of the energy  \(E\) and angular momentum \(L_{z}\). We can evaluate these quantities by equating their rate of change with the flux carried away by the GWs, \(\dot{E}\) and \(\dot{L_{z}}\). We will use the radiation fluxes derived by Gair \& Glampedakis~\cite{improved} augmented  with accurate BH perturbation theory (BHPT) results that include small body spin corrections~\cite{ tanaka}, i.e., 

\begin{eqnarray}
\dot{E} &=& -\frac{32}{5} \frac{\mu^2}{M }\left(\frac{1}{r}\right)^{5}\Bigg\{ 1 - \frac{1247}{336}\left(\frac{1}{r}\right)+ \left(4\pi-  \frac{73}{12}q-\frac{25}{4}\eta\chi \right)\left(\frac{1}{r}\right)^{3/2} + \left( - \frac{44711}{9072} + \frac{33}{16}q^2 + \frac{71}{8} q\eta\chi\right)\left(\frac{1}{r}\right)^2 \nonumber \\ && +\, \textrm{higher order Teukolsky fits}\Bigg\}, \nonumber\\
\dot{L}_z&=& -\frac{32}{5} \frac{\mu^2}{M} \left(\frac{1}{r}\right)^{7/2} \Bigg\{ 1 - \frac{1247}{336}\left(\frac{1}{r}\right)+ \left(4\pi-  \frac{61}{12}q-\frac{19}{4}\eta\chi \right)\left(\frac{1}{r}\right)^{3/2} + \left( - \frac{44711}{9072} + \frac{33}{16}q^2 + \frac{59}{8} q\eta \chi\right)\left(\frac{1}{r}\right)^2 \nonumber \\ && +\, \textrm{higher order Teukolsky fits}\Bigg\}.
\label{new_Ldot}
\end{eqnarray}

 \noindent The ``higher order Teukolsky fits'' are given in~\cite{improved}. We do not give these explicitly here, as they are not needed to derive the conservative corrections. However, we will include them to evolve orbits when we generate waveforms.

To evolve a circular orbit  for a spinless particle in a Kerr background, we need only to specify the angular momentum or the energy flux, as they are related by the `circular goes to circular' rule~\cite{ori}

\begin{equation}
\dot E_{\rm }(r, \chi\rightarrow 0) = \pm \frac{1}{r^{3/2} \pm q } \dot L_{z}(r,\chi\rightarrow 0 ) = \Omega_{\rm}(r,\chi\rightarrow 0 ) \dot L_{z}(r,\chi\rightarrow 0 ),
\label{cirkerr}
\end{equation}

\noindent where \( \mathrm{d} \phi / \mathrm{d} t = \Omega(r)\), is the azimuthal velocity of the orbit. We can use exactly the same scheme for spinning particles, as Tanaka et al. \cite{tanaka} showed that the assumption that a circular orbit remains circular under radiation reaction is consistent with the energy and angular momentum loss rates at linear order in the spin of the particle. Thus, Eq. \eqref{cirkerr} now reads \cite{mine, tanaka}

\begin{equation}
\dot E_{\rm }(r) = \pm \frac{1}{r^{3/2} \pm q }\left(1-\frac{3}{2}\eta\chi\frac{\pm \sqrt{r}-q}{r^2 \pm q \sqrt{r}}\right)  \dot L_{z}(r ).
\label{cirsp}
\end{equation}

Furthermore, the evolution in time of the radial coordinate is given by
\begin{equation}
\label{6}
\dot r= \frac{\mathrm{d} r}{\mathrm{d} E}\dot E= \frac{\mathrm{d} r}{\mathrm{d} L_z}\dot L_z .
\end{equation}
  \noindent In the non-spinning case, $\chi=0$, using the exact geodesic expression for  $\mathrm{d} r/\mathrm{d} L_z$ generates inspirals that are closer to Teukolsky based evolutions than expanding the above expression at 2PN order \cite{cons} and we will adopt this approach here as well. However, we will need the 2PN expression in the following, which is

 \begin{eqnarray}
\label{7}
\frac{\mathrm{d}r}{\mathrm{d}t} &=& - \frac{64}{5}\frac{\eta}{M} \left(\frac{1}{r}\right)^{3}\Bigg\{1- \frac{743}{336}\left(\frac{1}{r}\right) + \left(4 \pi - \frac{133}{12} q -\frac{35\, \eta\chi}{4}\right)\left(\frac{1}{r}\right)^{3/2} \nonumber\\ &+& \left(\frac{34103}{18144} + \frac{81}{16} q^{2} + \frac{95 \, q\eta\chi}{8}\right)\left(\frac{1}{r}\right)^{2}\Bigg\}.
\end{eqnarray}

\noindent Up to this point our analysis has been incomplete as we have only considered the three shape integrals of the motion. There are also three positional constants of the motion, which label the position of the test particle along the geodesic trajectory at some fiducial time. The self--interaction between the inspiralling body and the central BH has two main pieces. So far we have considered only the dissipative or radiative self--force, which drives the evolution of the shape constants. However, there is also a second piece, the conservative self--force, which affects the positional constants of the motion. Effectively, this changes the frequency of an orbit at a given radius but does not cause the orbit to evolve.

The conservative self--force has two parts: one is oscillatory and averages to zero, but there is also a secular piece which leads to accumulation of a phase error over time. In our model, we will include this effect by amending the evolution equation for the \(\phi\) frequency as follows,

\begin{equation}
\frac{\mathrm{d}\phi}{\mathrm{d}t} = \left(\frac{\mathrm{d}\phi}{\mathrm{d}t}\right)_{\mathrm{geo}}\bigg(1+  \delta \Omega \bigg).
\label{7.1}
\end{equation}

\noindent This equation includes the phase derivative for a geodesic, labeled by the subscript ``geo'' which is given implicitly in Eq.~(\ref{cirsp}), and a frequency shift which will depend on the instantaneous orbital parameters. A problem arises here because to compute the necessary frequency shifts within our framework, i.e., BHPT, would require self--force calculations. At present, Barack \& Sago have computed gravitational self--force corrections for particles moving on circular and eccentric bound geodesic orbits around a Schwarzschild BH in the Lorenz gauge \cite{sagoec, sago}. Warburton \& Barack have recently computed the self--force on a scalar charge for Kerr circular and eccentric  equatorial orbits in the same gauge \cite{warleor,war}. The extension of this work to Kerr eccentric inclined geodesic orbits is still a challenging endeavour mainly because there is not a formal framework to deal with Lorenz--gauge metric perturbations in the frequency domain --- the natural arena in which  these calculations are carried out.

However, we do know conservative corrections in the PN framework up to 2PN order which include spin--orbit, spin--spin couplings and finite mass contributions  \cite{blanchet}. We can combine these expressions with the radiative self--force obtained  by Tanaka et al. \cite{tanaka} based on the Teukolsky and Sasaki--Nakamura formalisms for perturbations around a Kerr BH, which includes terms of order \(q^{2},\, \eta\chi, \, q\eta\chi\). Using these results,  we will extend the method originally proposed by Babak et al. \cite{kludge}, who computed the 1PN conservative correction for circular orbits in the Schwarzschild space-time. This idea was subsequently extended by Huerta \& Gair \cite{cons}, who included conservative corrections in a kludge model at 2PN order for Kerr circular-equatorial orbits. In this paper we shall extend the latter model by including  small body spin effects at linear order, and their corresponding conservative corrections at 2PN order.

The idea is to correct the kludge expressed in a particular coordinate system, in order to ensure that asymptotic observables are consistent with PN results in the weak field. In particular, we aim to modify the orbital frequency and its first time derivative. By modifying these two quantities, we can both  identify coordinates between our kludge model and the PN formalism, and find the missing conservative pieces.  

From equation \eqref{cirsp}, and using \(\hat{s}=\eta\chi\), we find that at 2PN order the orbital frequency takes the form 

\begin{equation}
\label{eq.5}
\Omega = \frac{1}{M}\left(\frac{1}{r}\right)^{3/2}\left(1- \left(q + \frac{3}{2}\eta\chi\right)\left(\frac{1}{r}\right)^{3/2} + \frac{3}{2} q\eta\chi\left(\frac{1}{r}\right)^{2} + O\left(\frac{1}{r}\right)^{5/2}\right),
\end{equation}

\noindent which, following Eq.~\ref{7.1}, we now augment by including conservative corrections

\begin{eqnarray}
\frac{\mathrm{d}\phi}{\mathrm{d}t} \equiv \Omega &=& \frac{1}{M}\left(\frac{1}{r}\right)^{3/2}\left(1- \left(q + \frac{3}{2}\eta\chi\right)\left(\frac{1}{r}\right)^{3/2} + \frac{3}{2} q\eta\chi\left(\frac{1}{r}\right)^{2} \right)\bigg(1 + \delta \Omega \bigg),  \nonumber\\ 
 & =& \frac{1}{M}\left(\frac{1}{r}\right)^{3/2}\left(1- \left(q + \frac{3}{2}\eta\chi\right)\left(\frac{1}{r}\right)^{3/2} + \frac{3}{2} q\eta\chi\left(\frac{1}{r}\right)^{2} \right)\Bigg\{1 +  \nonumber\\ &+&\eta \left( d_0 + d_1 \left(\frac{1}{r}\right)+ (d_{1.5} + q\, f_{1.5} +\chi\, g_{1.5} )\left(\frac{1}{r}\right)^{3/2} + (d_2 + k_2\,q\chi)\left(\frac{1}{r}\right)^{2}\right)\Bigg\}.\label{omCCone}
\end{eqnarray}

 \noindent We will use this expansion for \(\Omega_{\rm geo}\) only to derive the conservative corrections. As with $\mathrm{d} r/\mathrm{d} L_z$, it has been shown that more reliable waveforms can be obtained by including the full geodesic frequency where it is known, see \eqref{7.1}, and this will be the approach used in section~\ref{s4}. It is inconsistent to include some pieces of the evolution at arbitrary PN order, while including the conservative corrections only at 2PN. However, the lower order effects that we are including with greater accuracy do have a more significant impact on the waveform~\cite{improved}.

The expansion in $r$ is an expansion in \(v^{2}= 1/r\). To derive the conservative corrections, we choose to leave the time derivative of the radial coordinate unchanged and given by equation \eqref{6}/\eqref{7}. This amounts to a choice of gauge in which the $\eta^2$ terms of ${\rm d}r/{\rm d}t$  that are not proportional to \(\chi\)  vanish. Differentiation of~(\ref{omCCone}) then gives \(\mathrm{d} \Omega/ \mathrm{d} t\) for the kludge, 

\begin{eqnarray}
\frac{\mathrm{d} \Omega}{ \mathrm{d} t} &=& \frac{96}{5}\frac{\eta}{M^2}\left(\frac{1}{r}\right)^{11/2}\Bigg\{ 1+ \eta\, d_0 + \frac{1}{r} \left(-\frac{743}{336} + \eta \left(\frac{5}{3} d_1 - \frac{743}{336} d_0 \right)\right) + \nonumber\\  &+& \left(\frac{1}{r}\right)^{3/2}\left(4 \pi - \frac{157}{12} q + \eta \left(4 \pi d_0 + 2 d_{1.5} + q \left(2 f_{1.5}- \frac{157}{12}d_0\right) +\chi\left(-\frac{47}{4} +2g_{1.5}\right) \right)\right) \nonumber \\ &+& \left(\frac{1}{r}\right)^{2}\left(\frac{34103}{18144}+ \frac{81}{16}q^2 +\eta\left(\frac{34103}{18144}d_0 + \frac{81}{16}q^2 d_0  - \frac{3715}{1008}d_1 + \frac{7}{3} d_2 + \left(\frac{123}{8}+\frac{7}{3}\,k_2\right)q\chi\right)\right) \Bigg\}.\nonumber\\
\label{odot}
\end{eqnarray}

Note we have used \(\hat{s}= \eta \chi\), with \(\chi = S_1/\mu^2\). We may now write down a coordinate transformation  to relate our coordinates with those used in the PN formalism, namely,
\begin{eqnarray}
r &=& \frac{R}{M} \Bigg\{ 1 + \left(\frac{M}{R}\right)b_1 + \left(\frac{M}{R}\right)^{3/2}(b_{1.5} + q \,w_{1.5} + \eta \chi\,v_{1.5}) + \left(\frac{M}{R}\right)^{2}(b_2 + b_{2.1} q\,\eta\chi)\nonumber\\  &+& \eta \left( c_0 +\left(\frac{M}{R}\right)c_1 + \left(\frac{M}{R}\right)^{3/2}(c_{1.5} + q\, l_{1.5}+ \gamma \chi) + \left(\frac{M}{R}\right)^{2}\left( c_2+ c_{2.1} q\,\chi\right)\right)\Bigg\}, \label{coord}
\end{eqnarray}

\noindent where \(R\) denotes the PN semi-major axis. We can now substitute this expression for the coordinate transformation into relations \eqref{omCCone} and \eqref{odot}. 

The final stage of the computation is to compare the expressions for \(\Omega\) and \(\dot \Omega\), where a dot denotes \(\mathrm{d}/ \mathrm{d} t\),  with the available PN expansions. The PN expansions are available to higher order in the mass ratio \(\eta\), but we keep \(\eta\) only to the same order  as the kludge, Eq. \eqref{odot}. The PN expressions for the orbital frequency and its first time derivative are given by \cite{blanchet}

\begin{eqnarray}
\Omega_{PN}^{2} &=& \frac{m_{\textrm{T}}}{R^{3}}\Bigg\{1- \frac{m_{\textrm{T}}}{R}\left(3 - \eta \right) - \left( \frac{m_{\textrm{T}}}{R}\right)^{3/2}\sum_{i}\left(2 \left(\frac{m_i}{m_{\textrm{T}}}\right)^{2} + 3 \eta \right) \boldsymbol{\hat{L} \cdot \chi_i} \nonumber\\ &+& \left( \frac{m_{\textrm{T}}}{R}\right)^{2} \left( 6 + \frac{41}{4} \eta - \frac{3\eta}{2}\left(\boldsymbol{\chi_1\cdot \chi_2} - 3\boldsymbol{\hat{L} \cdot \chi_1 \hat{L} \cdot \chi_2} \right)\right)\Bigg\},
\label{1}  
 \end{eqnarray}

\noindent where \(m_{\textrm{T}} = M + m \), \( \boldsymbol{\hat L}\) is a unit vector directed along the orbital momentum, \(\boldsymbol{\chi=\chi_1= S_1}/\mu^2\), \(\boldsymbol{q=\chi_2= S_2}/M^2\). Additionally \cite{blanchet},

\begin{eqnarray}
\dot \Omega_{PN} &=& \frac{96}{5}\eta m_{\textrm{T}}^{5/3}\omega^{11/3} \Bigg\{ 1- \left( \frac{743}{336} + \frac{11}{4} \eta \right)(m_{\textrm{T}} \omega)^{2/3} + (4 \pi - \beta)(m_{\textrm{T}}\omega) \nonumber \\ &+& \left( \frac{34103}{18144} + \frac{81}{16} q^{2} + \sigma + \eta \left( \frac{13661}{2016} + \zeta q^{2}\right) \right) (m_{\textrm{T}} \omega)^{4/3} \Bigg\},
\label{2}
\end{eqnarray}

\noindent where  the constant \(\zeta\) was determined in \cite{cons}, and has the value \(\zeta = - 243/32\). This term guarantees that the PN framework and the perturbative approach coincide in the test mass particle limit \(\eta \rightarrow 0\). The spin--orbit \(\beta\) and spin--spin parameters \(\sigma\) are given by 

\begin{eqnarray}
\beta &=& \frac{1}{12} \sum_{i}\left( 113 \frac{m^2_{i}}{ m_{\textrm{T}}^2} + 75 \eta \right) \boldsymbol{\hat L \cdot \chi_i}, \nonumber\\
\sigma&=& \frac{\eta}{48}\left(-247 \boldsymbol{\chi_1\cdot \chi_2} +721\boldsymbol{\hat{L} \cdot \chi_1 \hat{L} \cdot \chi_2}\right).
\end{eqnarray}

\noindent To be consistent with the perturbative approach outlined above, we shall assume that the spin of the small particle is perpendicular to the equatorial plane (this guarantees that the orbit remains circular-equatorial), and parallel to the momentum of the central Kerr BH. For prograde orbits, the spin and spin--spin corrections will play a more significant role, and so we will focus our analysis on those. We now rewrite these expressions in a convenient way to take the small mass--ratio limit, by writing \(m_{\textrm{T}} = M (1+ \eta)\), giving

\begin{eqnarray}
\label{3} 
\Omega_{PN} &=& \frac{1}{M}\left(\frac{M}{R}\right)^{3/2}\Bigg\{1 + \frac{\eta}{2}- \frac{M}{R}\left(\frac{3}{2} + \frac{7}{4} \eta \right) -  \left( \frac{M}{R}\right)^{3/2}\left(q+ \eta\left(\frac{3}{2}q + \frac{3}{2}\chi \right) \right) \nonumber \\ &+& \left( \frac{M}{R}\right)^{2} \left( \frac{15}{8}  + \eta\left(\frac{169}{16}+ \frac{3}{2}q\,\chi \right)\right)\Bigg\},
\end{eqnarray}

\begin{eqnarray}
\label{4}
\dot \Omega_{PN} &=& \frac{96}{5}\frac{\eta}{M^2} \left(\frac{M}{R}\right)^{11/2} \Bigg\{ 1 + \frac{3}{2}\eta- \left( \frac{2591}{336} + \frac{13571}{672} \eta \right)\frac{M}{R} \nonumber \\ &+& \left(4 \pi - \frac{157}{12}q + \eta \left(12 \pi - \frac{149}{6}q-\frac{47}{4}\chi\right)\right)\left(\frac{M}{R}\right)^{3/2} \nonumber \\ &+& \left( \frac{22115}{648} + \frac{81}{16} q^{2}  + \eta \left( \frac{87044}{567} + \frac{81}{8}q^{2}+ \frac{123}{8}q\,\chi \right)\right) \left(\frac{M}{R}\right)^{2} \Bigg\}.\nonumber\\ 
\end{eqnarray}

\noindent A direct comparison between the expressions for the orbital frequencies and their first time derivatives allow us to solve simultaneously for the various coefficients of Eqs. \eqref{omCCone}, \eqref{odot} and \eqref{coord}. We find that the non--vanishing parameters are

\begin{eqnarray}
b_1 = 1, \qquad c_0 =- \frac{1}{4}, \qquad c_1 = \frac{845}{448}, \qquad d_0 = \frac{1}{8}, \qquad d_1 = \frac{1975}{896} \nonumber\\
c_{1.5}= -\frac{9}{5}\pi, \qquad d_{1.5}= -\frac{27}{10} \pi, \qquad f_{1.5}= -\frac{191}{160}, \qquad l_{1.5} = -\frac{91}{240}\nonumber\\
c_2 = -\frac{2 065 193}{677 376},  \qquad  d_2 = \frac{1 152 343}{451 584}.
\label{10}
\end{eqnarray}

\noindent Having found the required corrections, we can now explore whether small body spin effects are important for signal detection. We will do this in Section~\ref{s5}, by computing the inspiral trajectory using the relations \eqref{new_Ldot}, \eqref{6}, and using the azimuthal frequency amended with conservative corrections, i.e., 

\begin{eqnarray}
\Omega= &&\frac{1}{r^{3/2} + q }\left(1-\frac{3}{2}\eta\chi\frac{ \sqrt{r}-q}{r^2 + q \sqrt{r}}\right)\Bigg\{1 +  \nonumber\\&&\eta\left( d_0 + d_1 \left(\frac{1}{r}\right)+ (d_{1.5} + q\, f_{1.5} +\chi\, g_{1.5} )\left(\frac{1}{r}\right)^{3/2} + (d_2 + k_2\,q\chi)\left(\frac{1}{r}\right)^{2}\right)\Bigg\}.
\label{omegacc}
\end{eqnarray}

\noindent We note that this equation includes corrections up to 2PN order, characterised by the coefficients $d_0$ etc.. In Section~\ref{s5}, we will assess the importance of these corrections for signal detection and parameter estimation, by turning these parameters on and off (i.e., setting a subset of them to zero) in the model.

Previous studies have ignored small body spin effects, in particular for EMRIs, since they are expected to be small for mass rations \(\eta \sim 10^{-5}\). However, even though an accurate measurement of the spin of the small body will be unlikely for these systems, we will discuss whether this additional parameter has a bearing on the accuracy with which other parameters can be determined. Furthermore, we will study various systems to find the mass ratio threshold at which the small spin parameter starts to be measurable through GW observations, and will also explore whether small object spin conservative corrections are relevant for parameter estimation and detection. 

The following section briefly outlines the waveform model to be used in these studies.

\section{Waveform model}
\label{s3}
Having derived an accurate scheme for the inspiral evolution, we need a prescription for the waveform model. We shall generate our waveforms using a flat--space quadrupole wave generation formula applied to the trajectory of the inspiralling object in Boyer-Lindquist coordinates, which we identify with spherical--polar coordinates in a flat--space. Using this ``pseudo--flat space'' quadrupole moment approximation is not correct, but this prescription has been seen to work well in practice, as shown in \cite{kludge}.

In the transverse--traceless (TT) gauge, the metric perturbation \(h_{i j}(t)\) is given by,
\begin{equation}
h_{i j}(t) = \frac{2}{D}\left(P_{ik}P_{jl}- \frac{1}{2}P_{ij}P_{kl}\right)\ddot  I^{kl},
\label{13}
\end{equation}
\noindent where \(D\) is the distance to the source, \(\eta_{ij}\) is the flat Minkowski metric, the  projection operator \(P_{ij} = \eta_{ij} - \hat n_{i}\hat n_{j}\), \(\hat n_{i}\) is a unit vector in the direction of propagation, and \(\ddot  I^{kl}\) is the second time derivative of the inertia tensor \cite{mtw}. In the EMRI framework this takes the form \(  I^{kl}= m r^i(t)r^j(t)\),   where \(r^i(t)\) represents the position vector of the CO with respect to the MBH in the pseudo--flat space. 

\subsection{Implementation of LISA's response function}

Since most of the SNR of the binary inspirals considered in these studies accumulates at frequencies \(f\lesssim 10\)mHz, it should be adequate to use the low-frequency approximation to the detector response. Following \cite{cutlerold}, the LISA response in this regime may be written as

\begin{equation}
h_{\alpha}(t)= \frac{\sqrt{3}}{2 D} \Big[F_\alpha ^ +(t)A^ +(t) + F_\alpha ^ \times(t) A^ \times(t)\Big],
\label{14}
\end{equation}
\noindent where $\alpha = I,II$ refers to the two independent Michelson-like detectors that constitute the LISA response at low frequencies. The functions \(A^ {+\, , \times} (t)\) are the polarization coefficients given by
\begin{equation}
A ^ + = -a_+ [1+ (\hat{a} \cdot \hat n)^2], \qquad A^ \times = 2 a_\times (\hat a \cdot \hat n),
\label{15}
\end{equation}
\noindent where \(\hat a\) is a unit vector along the SMBH's spin direction, and \(a_+,a_\times\) are given by 
\begin{displaymath}
a_+= \frac{1}{2}\left( \ddot  I^{11}- \ddot  I^{22}\right), \qquad a_\times= \ddot  I^{12}.
\end{displaymath}
\noindent The antenna pattern functions \(F_\alpha ^{+ \times}\) are given by
\begin{eqnarray}  
F^{+}_{I} &=&  \frac{1}{2}(1+\cos^2\theta)\cos(2\phi)\cos(2\psi)
-\cos\theta\sin(2\phi)\sin(2\psi), \nonumber\\
F^{\times}_{I} &=& \frac{1}{2}(1+\cos^2\theta)\cos(2\phi)\sin(2\psi)
+\cos\theta\sin(2\phi)\cos(2\psi),
\label{16} \\
F^{+}_{II} &=&  \frac{1}{2}(1+\cos^2\theta)\sin(2\phi)\cos(2\psi)
+\cos\theta\cos(2\phi)\sin(2\psi), \nonumber\\
F^{\times}_{II} &=& \frac{1}{2}(1+\cos^2\theta)\sin(2\phi)\sin(2\psi)
-\cos\theta\cos(2\phi)\cos(2\psi).
\label{17}
\end{eqnarray}  

\noindent The various angles in the previous expressions represent the source's sky location in a detector based coordinate system, (\(\theta,\phi\)), and the polarization angle of the wavefront, \(\psi\). These can be re--written in a fixed, ecliptic--based coordinate system. If we denote the source co--latitude and azimuth angles and the direction of \(\hat a\) in this fixed coordinate system by (\(\theta_S,\phi_S\)) and (\(\theta_K,\phi_K\)) respectively, then 
\begin{eqnarray} 
\cos\theta(t) &=& \frac{1}{2}\cos\theta_S-\frac{\sqrt{3}}{2}\sin\theta_S
\cos[\bar\phi_0+2\pi(t/T)-\phi_S], \nonumber\\
\phi(t) &=& \bar\alpha_0+2\pi(t/T)+ \tan^{-1}\Bigg\{
\frac{\sqrt{3}\cos\theta_S+\sin\theta_S\cos[\bar\phi_0+2\pi(t/T)-\phi_S]}
{2\sin\theta_S\sin[\bar\phi_0+2\pi(t/T)-\phi_S]}\Bigg\},\nonumber\\
\tan\psi & = & \Bigg\{\frac{1}{2}\cos\theta_K-\frac{\sqrt{3}}{2}
\sin\theta_K \cos[\bar\phi_0+2\pi(t/T)-\phi_K] \nonumber \\ &-& \cos\theta(t)\left[
\cos\theta_K\cos\theta_S+\sin\theta_K\sin\theta_S\cos(\phi_K-\phi_S)\right] \Bigg\}\Big /  
 \nonumber\\ && \Bigg\{
\frac{1}{2}\sin\theta_K\sin\theta_S\sin(\phi_K-\phi_S)
-\frac{\sqrt{3}}{2}\cos(\bar\phi_0+2\pi t/T)\nonumber\\ &&\{\cos\theta_K\sin\theta_S\sin\phi_S
-\cos\theta_S\sin\theta_K\sin\phi_K\} \nonumber\\ &-&
\frac{\sqrt{3}}{2}\sin(\bar\phi_0+2\pi t/T) \left(\cos\theta_S\sin\theta_K\cos\phi_K
-\cos\theta_K\sin\theta_S\cos\phi_S\right)\Bigg\},\label{18}
\end{eqnarray}
\noindent where \( \bar\phi_0, \bar\alpha_0\) are constant angles which represent the orbital and rotational phase of the detector at \(t=0\). We will set both of these to zero in our analysis. Additionally, \(T\) is the orbital period, which is 1 year. Barack and Cutler \cite{cutler} write these expressions in terms of  \(\theta_L, \phi_L\) which specify the direction of the CO's orbital angular momentum in the ecliptic--based system. In this case the orbits under consideration are circular and equatorial, so the angular momentum vector of the orbiting body does not precess about the SMBH spin \(\hat{a}\), and hence \(\theta_K =\theta_L , \phi_K = \phi_L\).

The last ingredient in the detector response is the Doppler phase modulation. If \(\Phi(t)\) denotes the phase of the waveform, the inclusion of the Doppler modulation shifts the phase as follows \cite{cutler}
\begin{equation}
\Phi(t)\to \Phi(t)+ 2 \frac{\mathrm{d}\phi}{\mathrm{d}t} R \sin\theta_S \cos[2\pi(t/T)-\phi_S],
\label{20}
\end{equation}
\noindent where  \(R= 1 \textrm{AU/c}= 499.00478 \textrm{s}\) and \(\mathrm{d}\phi/\mathrm{d}t\) is the azimuthal velocity of the orbit of the inspiralling object (see Eq. \eqref{omCCone}).

\section{Noise induced parameter errors}
\label{s4}
This section contains an overview of the basic elements of signal analysis that we will use in Section~\ref{s5} to estimate the accuracies with which GW observations will be able to determine the system parameters.

The measured strain \(s(t)\) in a GW detector is a time series that contains both a true GW signal, \(h(t)\), and instrumental noise, \(n(t)\). In the context of LISA, the output of the equivalent two arm Michelson detectors can be represented as
 
\begin{equation}
s_{\alpha}(t) = h_{\alpha}(t) + n_{\alpha}(t), \qquad \alpha = \textrm{I, \,II}.
\label{21}
\end{equation}

\noindent Given two time series, we define the overlap as 

\begin{equation}\label{23}
\left( {\bf p} \,|\, {\bf q} \right)
\equiv 2\sum_{\alpha} \int_0^{\infty}\left[ \tilde p_\alpha^*(f)
\tilde q_\alpha(f) + \tilde p_\alpha(f) \tilde q_\alpha^*(f)\right]
/S_n(f)\,\mathrm{d}f,
\end{equation}

\noindent where \(S_n(f)\) stands for the one-sided spectral density of the instrumental noise. Assuming that each Fourier component  of the noise, \({\tilde n_{\alpha}}(f)\), is Gaussian distributed, and uncorrelated with other Fourier components (i.e., the noise is stationary), the ensemble average of the Fourier components of the noise has the property 

\begin{equation}
\label{22}
\langle {\tilde n_\alpha}(f) \, {\tilde n_\beta}(f^\prime)^* \rangle =\frac{1}{2}
\delta(f - f^\prime) S_n(f) \delta_{\alpha \beta}.
\end{equation}

\noindent The former relation defines the spectral density  \(S_n(f)\), which is the same for the two Michelson data streams.  Furthermore, the probability distribution function for the noise \(n(t)\) is given by 

\begin{equation}
\label{24}
p({\bf n=n}_0) \propto \textrm{exp}\left(- \frac{\left( {\bf n}_0\, |\,{\bf n}_0 \right)}{2}\right).
\end{equation}
\noindent  We can interpret Eq. \eqref{24} as the probability that the actual noise realization is \({\bf n}_0\). Assuming that we have made a detection,  i.e., the output of the detector is given by \({\bf s}(t)= {\bf h}(t;\theta_{\textrm{true}}) + {\bf n}_0(t)\), where \({\bf n_0(t)}\) is the specific realization of the noise and \(\theta_{\textrm{true}}\) is the unknown true value of the parameters of the source,  the likelihood of measuring the output signal is

\begin{equation}
\Lambda\left( {\bf s}\,|\, \theta_{\textrm{true}} \right) \propto \textrm{exp}\left( -\frac{\left({\bf s}- {\bf h}(\theta_{\textrm{true}} )\right)\, |\,\left({\bf s}- {\bf h}(\theta_{\textrm{true}})\right)}{2}  \right).
\label{lik}
\end{equation}

\noindent In order to reconstruct the most probable value of the parameters of the source, and compute their respective errors we need to calculate the probability of the parameters  given the data, i.e., the posterior probability, which, according to Bayes's theorem, is given by the product of the likelihood function, Eq. \eqref{lik}, and the prior probability  \(p_0(\theta_{\textrm{true}})\), i.e., 

\begin{equation}
\label{25}
p\left( {\bf h}(\theta_{\textrm{true}}) \,|\, {\bf s} \right) = N\,p_0(\theta_{\textrm{true}}) \textrm{exp}\left( \left( {\bf h}(\theta_{\textrm{true}})\, |\, {\bf s} \right)- \frac{1}{2}\left( {\bf h}(\theta_{\textrm{true}})\, |\, {\bf h}(\theta_{\textrm{true}}) \right)\right),
\end{equation} 

\noindent where the factor \(\left({\bf s}\, |\, {\bf s}\right)/2)\) has been absorbed into the normalization factor \(N\). For a given measured signal \({\bf s(t)}\), the gravitational waveform \({\bf h(t)}\) that best fits the data is the one that minimizes the quantity \(\left( {\bf s-h}\,|\, {\bf s-h} \right) \). This condition is also satisfied by the maximum likelihood estimators of the parameters, and corresponds to the point in parameter space with the highest  SNR in a matched filtering search, namely,

\begin{equation}
\frac{S}{N}[h(\theta^i)]= \frac{\left( {\bf s}\,|\, {\bf h} \right)}{\sqrt{\left( {\bf h}\,|\, {\bf h} \right)}}.
\label{26}
\end{equation}

\noindent In the limit of high SNR, a locally flat prior would be a 
reasonable assumption because the best--fit parameters will have a Gaussian distribution centered on the correct values. Hence, we can expand Eq.~\eqref{25} about the peak, $\hat{\theta}=\theta_{\textrm{true}} $, by setting $\theta^i= \hat{\theta^i} + \Delta \theta^i$, and find 

\begin{equation}
\label{27}
p(\Delta\theta \, |\,s)=\,{\cal N} \, e^{-\frac{1}{2}\Gamma_{ij}\Delta \theta^i
\Delta \theta^j}, \qquad \Gamma_{ij} \equiv \bigg( \frac{\partial {\bf h}}{ \partial \lambda^i}\, \bigg| \,
\frac{\partial {\bf h}}{ \partial \lambda^j }\bigg)_{|\theta=\hat{\theta}},
\end{equation}

\noindent where \( p(\Delta\theta \, |\,s)\) is the Gaussian probability distribution of the parameter estimation errors \( \Delta\theta \), and $\Gamma_{ij}$ is the Fisher Information Matrix. Additionally, for large SNR, the covariance of the posterior probability distribution, \((\Gamma^{-1})^{ij}\), gives the expectation value of the errors \(\Delta \theta^i\)

\begin{equation}
\label{29}
\left< {\Delta \theta^i} {\Delta \theta^j}
 \right>  = (\Gamma^{-1})^{ij} + {\cal O}({\rm SNR})^{-1} .
\end{equation}

\noindent In our analysis, we will use a simplified definition of the 
inner product, Eq. \eqref{23}. For white noise, i.e., \(S_n(f)= \textrm{const.}\), it  takes the simple form  \(2 S_n^{-1}\sum_{\alpha}\int_{-\infty}^{\infty} \, p_\alpha(t) q_\alpha(t) dt\), by Parseval's theorem \cite{cutler}. Following Barack \& Cutler \cite{cutler},  we can define the ``noise--weighted'' waveform as follows
\begin{equation}\label{30}
\hat h_{\alpha}(t) \equiv   \frac{h_{\alpha}(t)}{\sqrt{S_h\bigl(f(t)\bigr)}}, \qquad f(t) = \frac{1}{\pi}\frac{\mathrm{d}\phi}{\mathrm{d}t},
\end{equation}
\noindent and rewrite the Fisher matrix approximately as 
\begin{equation}\label{31}
\Gamma_{ab} = 2\sum_{\alpha}\int_0^T{\partial_a \hat h_{\alpha}(t) \partial_b \hat h_{\alpha}(t) \mathrm{d}t} \, .
\end{equation}

\subsection{Noise model}
The function \(S_h\bigl(f\bigr)\) is the total LISA noise, which has three components: instrumental noise, confusion noise from short--period galactic binaries, and confusion noise from extragalactic binaries. We use the same prescription as in Barack and Cutler~\cite{cutler}, namely
\begin{eqnarray}
S_h\bigl(f(t)\bigr)& =& {\rm min}\big\{S_h^{\rm inst}(f)/\exp(-\kappa T_{\rm mission}^{-1} \mathrm{d}N/\mathrm{d}f) + S_h^{\rm ex gal}(f) ,\nonumber\\ &&  S_h^{\rm inst}(f) + S_h^{\rm gal}(f)+  S_h^{\rm ex gal}(f)\big\}\, ,
\label{32}
\end{eqnarray}
\noindent where the various components have the following analytic forms \cite{cutler}
\begin{eqnarray}
S^{\rm inst}_h(f) &=& 9.18 \times 10^{-52}f^{-4} + 1.59 \times 10^{-41}
+ 9.18 \times 10^{-38}f^{2}\;\; {\rm Hz}^{-1}, \nonumber \\ 
S^{\rm gal}_h(f) &=& 2.1\times10^{-45}\,\left(\frac{f}{1{\rm Hz}}\right)^{-7/3} \; {\rm Hz}^{-1}, \nonumber \\ 
S_h^{\rm ex.\ gal} &=&
4.2 \times 10^{-47} \left(\frac{f}{1{\rm Hz}}\right)^{-7/3}\; {\rm Hz}^{-1}.
\label{33}
\end{eqnarray}
\noindent Here \(\mathrm{d}N/\mathrm{d}f\) is the number density of galactic white dwarf binaries per unit GW frequency, and \(\kappa\) is the average number of frequency bins that are lost when each galactic binary is fitted out. We use
\begin{equation}\label{34}
\frac{\mathrm{d}N}{\mathrm{d}f} = 2\times10^{-3}\,{\rm Hz}^{-1}\left(\frac{1\,{\rm Hz}}{f}\right)^{11/3}, \qquad \kappa T_{\rm mission}^{-1} = 1.5/{\rm yr},
\end{equation}
\noindent with \( T_{\rm mission} = 3\) yr, and \(\kappa =4.5\).

\section{Parameter estimation error results}
\label{s5}

In this Section we explore the accuracy with which LISA observations will be able to measure the spin of the inspiralling object. To effectively address this problem, we will estimate noise--induced errors for a variety of binary systems using the formalism described in Sections~\ref{s2} and \ref{s4}. To estimate the noise--induced errors we will use the inverse Fisher Matrix (see Eq.~\eqref{29}). We have verified that the Fisher Matrices used in our studies were convergent over several orders of magnitude in the offset used to compute the numerical waveform derivatives. The same convergent behaviour was exhibited by the associated inverse matrices. 

The noise--induced errors will be quoted for fixed values of the intrinsic parameters of the source, but with a Monte Carlo simulation over possible values of the extrinsic parameters. We compute the Fisher Matrix for a source at $D=1$Gpc, and the corresponding SNR using the expression
\begin{equation}\label{35}
{\rm SNR}^2= 2\sum_{\alpha=I,II}\int_{t_{\rm init}}^{t_{\rm LSO}}
\hat h_{\alpha}^2(t)dt.
\end{equation}
\noindent  We then renormalise the results to a fixed SNR, which was chosen separately for each of the systems considered. This choice was based on a Monte Carlo simulation of the SNRs, as described in the next subsection.

\subsection{Determination of typical SNRs}

To estimate a suitable ``typical'' SNR for each type of source to use as a reference, we carried out a Monte Carlo simulation in which the extrinsic parameters of each source were chosen randomly, with the events distributed uniformly in comoving volume out to a redshift of $z=1$, and detected in a certain time window at the detector, which we took to be one year. For each event, we computed the SNR and then looked at how the SNRs of the detected events were distributed. For systems involving the more massive inspiraling objects, events at redshifts $z>1$ could still be detected by LISA. However, given the uncertainties that exist in the formation mechanisms for IMBHs and the inspiral rates, we chose to assume events that occur only for redshifts $z \lesssim 1$. This would be a conservative assumption if it were used to compute the event rate for detected inspirals. However, if there are a large number of inspirals occurring at higher redshift, the reference SNRs we compute in this way could be somewhat larger than those of the systems LISA eventually detects. A full assessment of the parameter estimation accuracies that LISA will achieve would require detailed modelling of the source population that is beyond the scope of the current paper. The results presented here can be readily rescaled to other SNRs for any future study of this kind.

\begin{figure*}[ht]
\centerline{
\includegraphics[height=0.33\textwidth,  clip]{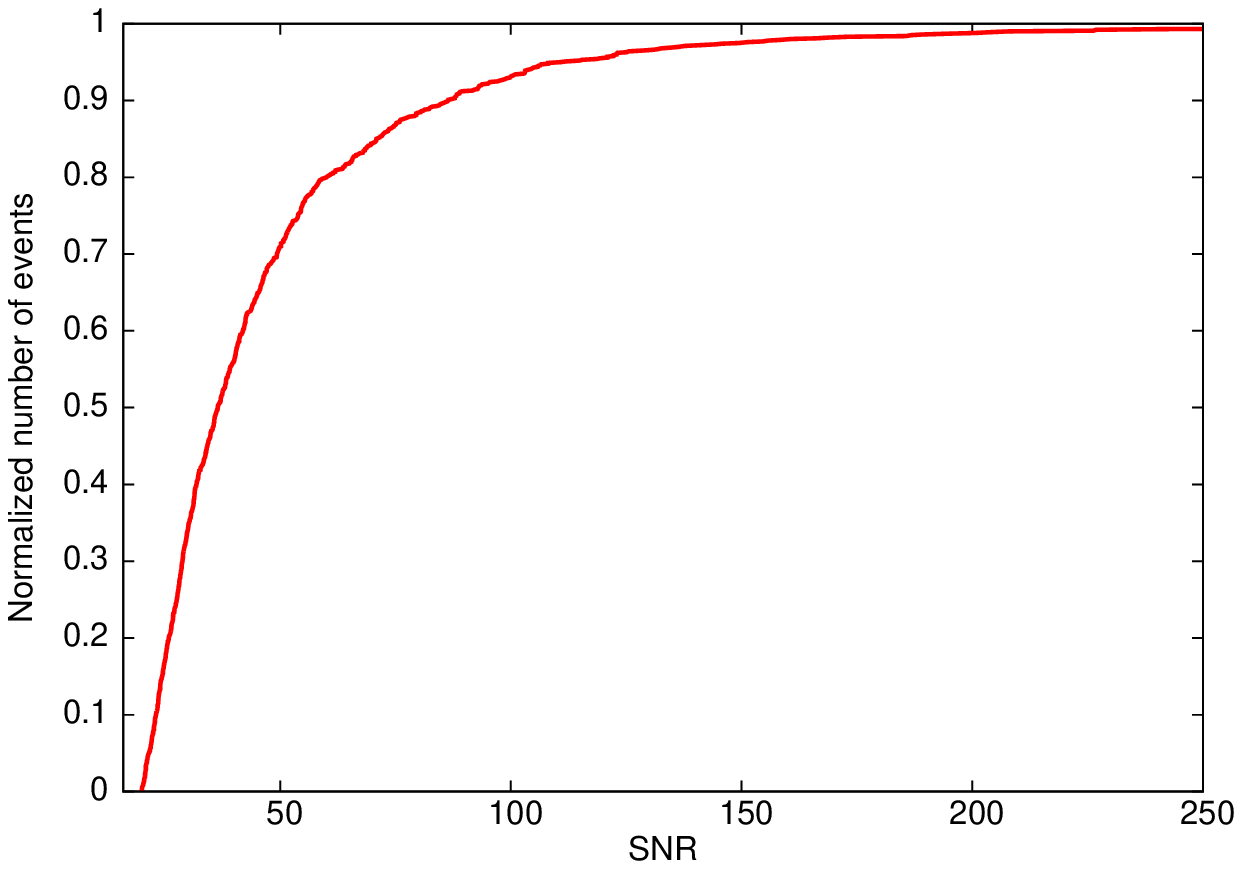}
\includegraphics[height=0.33\textwidth,  clip]{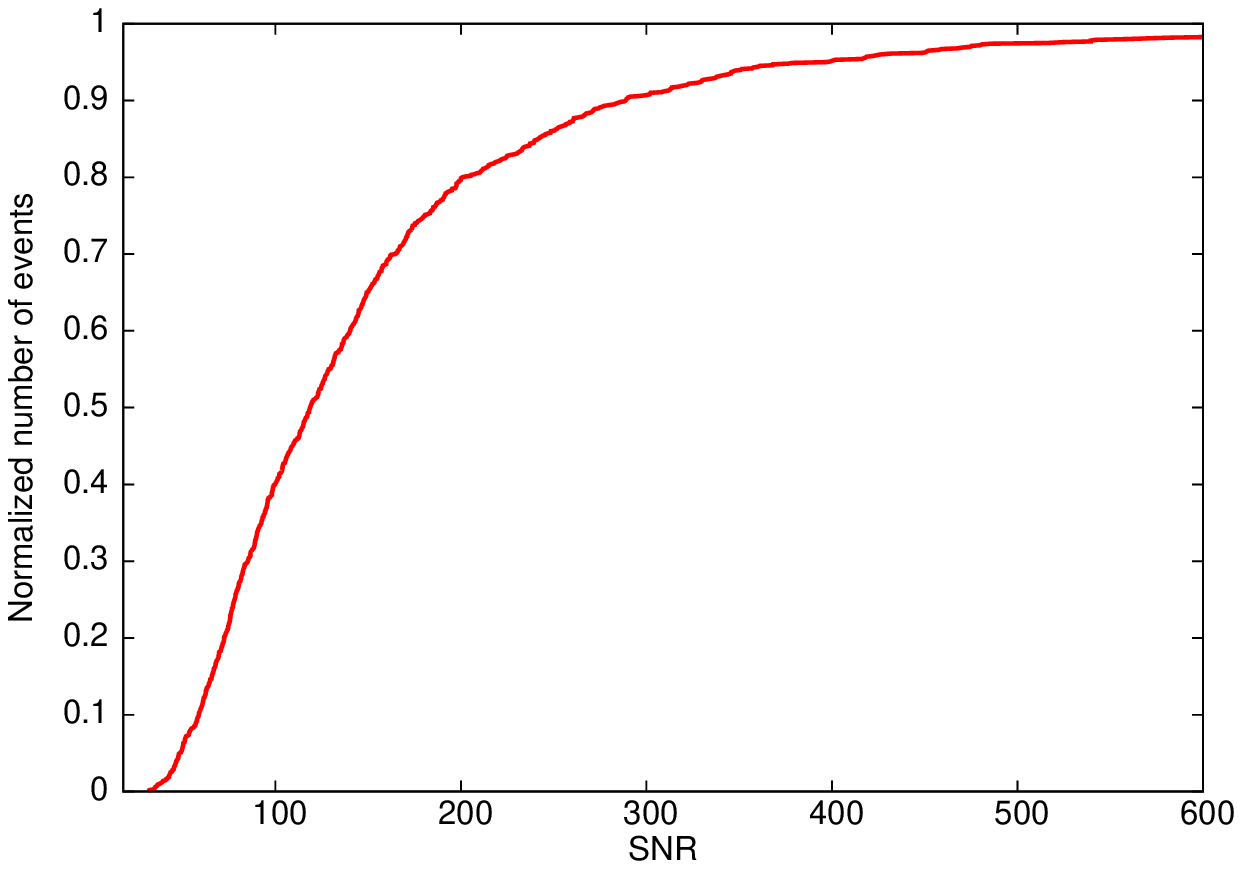}
}
\centerline{
\includegraphics[height=0.33\textwidth,  clip]{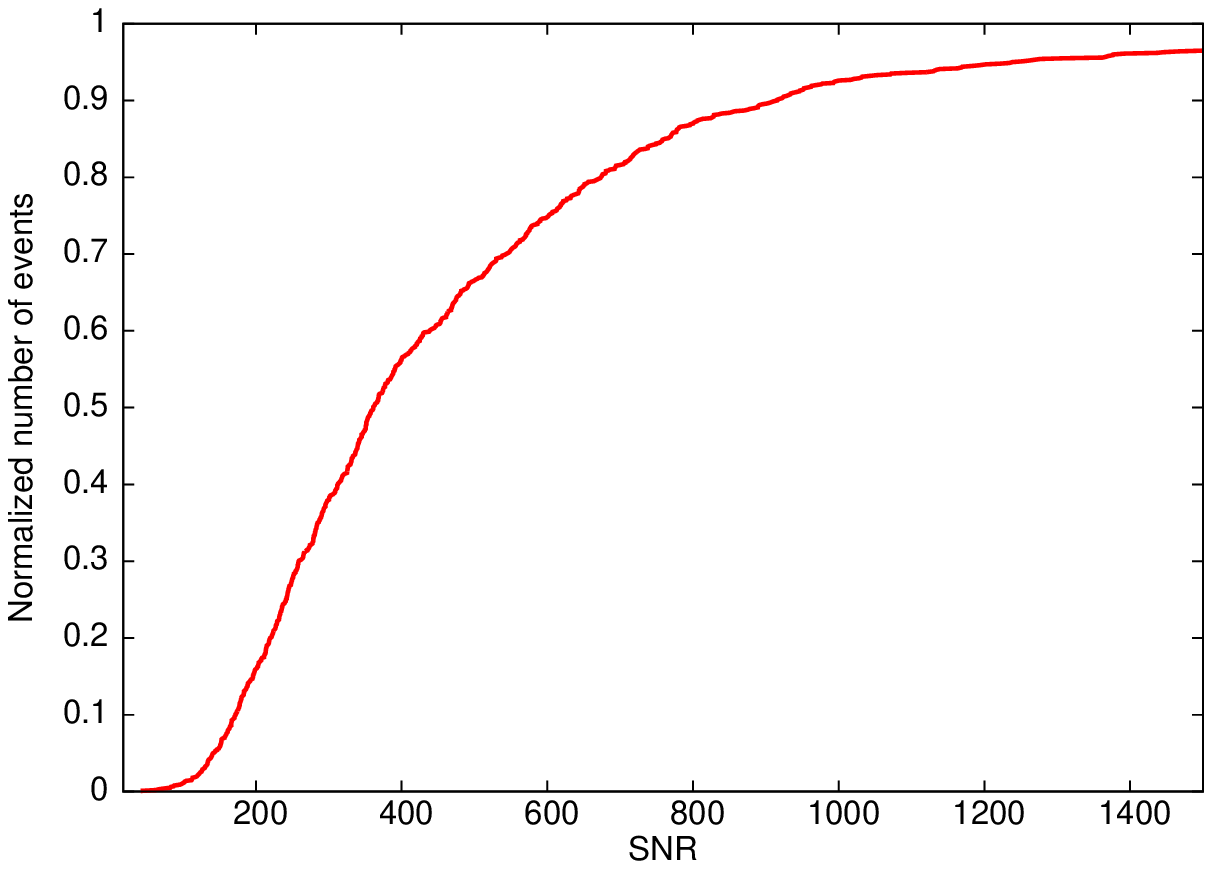}
\includegraphics[height=0.33\textwidth,  clip]{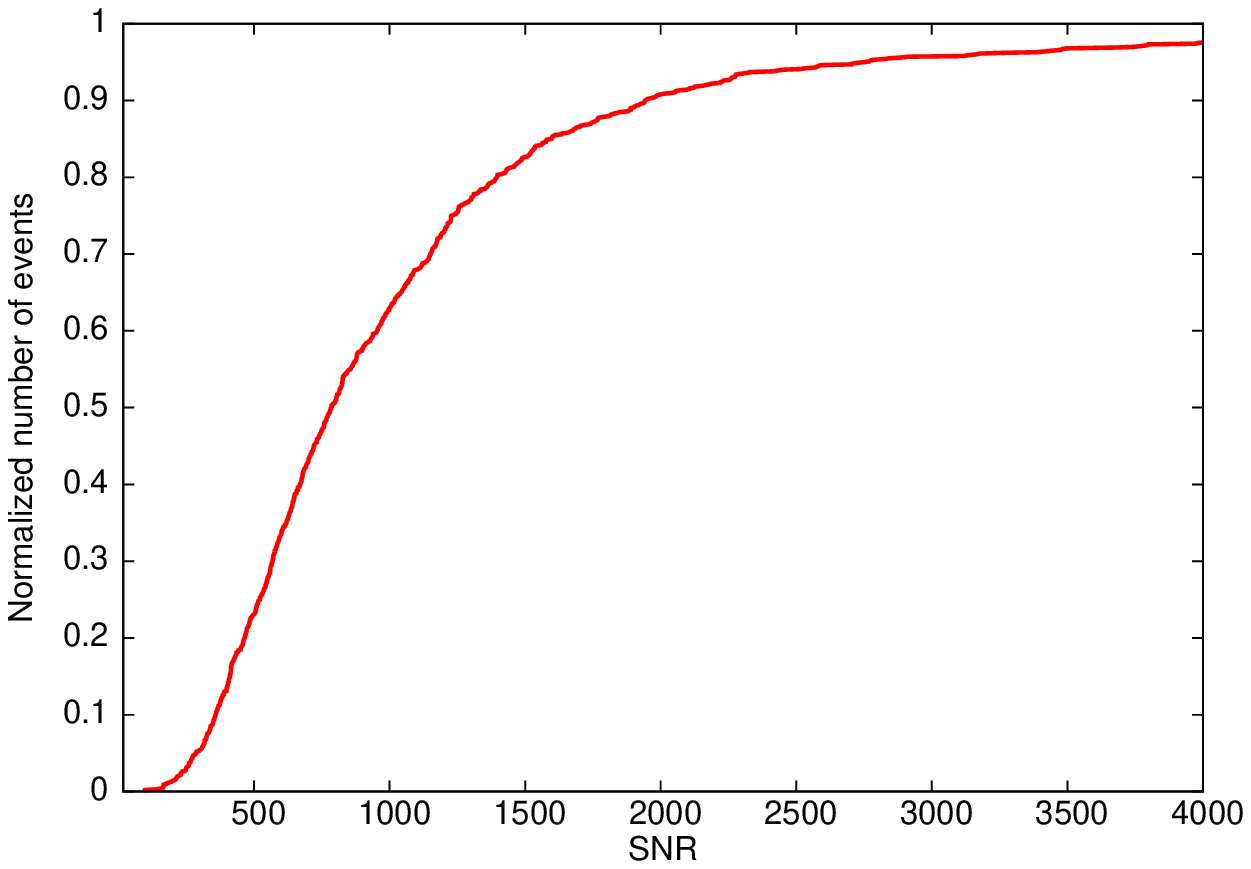}
}
\caption{The normalized cumulative distribution function for the signal--to--noise ratio of a cosmological population of binary systems with central BHs of redshifted mass  $M=10^6M_{\odot}$, and spin parameter $q=0.9$. The inspiralling BHs have specific spin parameter \(\chi =0.9\), and redshifted masses $\mu=10M_{\odot}$ (top left panel), $\mu=10^2M_{\odot}$ (top right), $\mu=10^3M_{\odot}$ (bottom left), and $\mu=5\times10^3M_{\odot}$ (bottom right).}
\label{smoothdis}
\end{figure*}

\begin{figure*}[ht]
\centerline{
  \includegraphics[height=0.33\textwidth,angle=0,  clip]{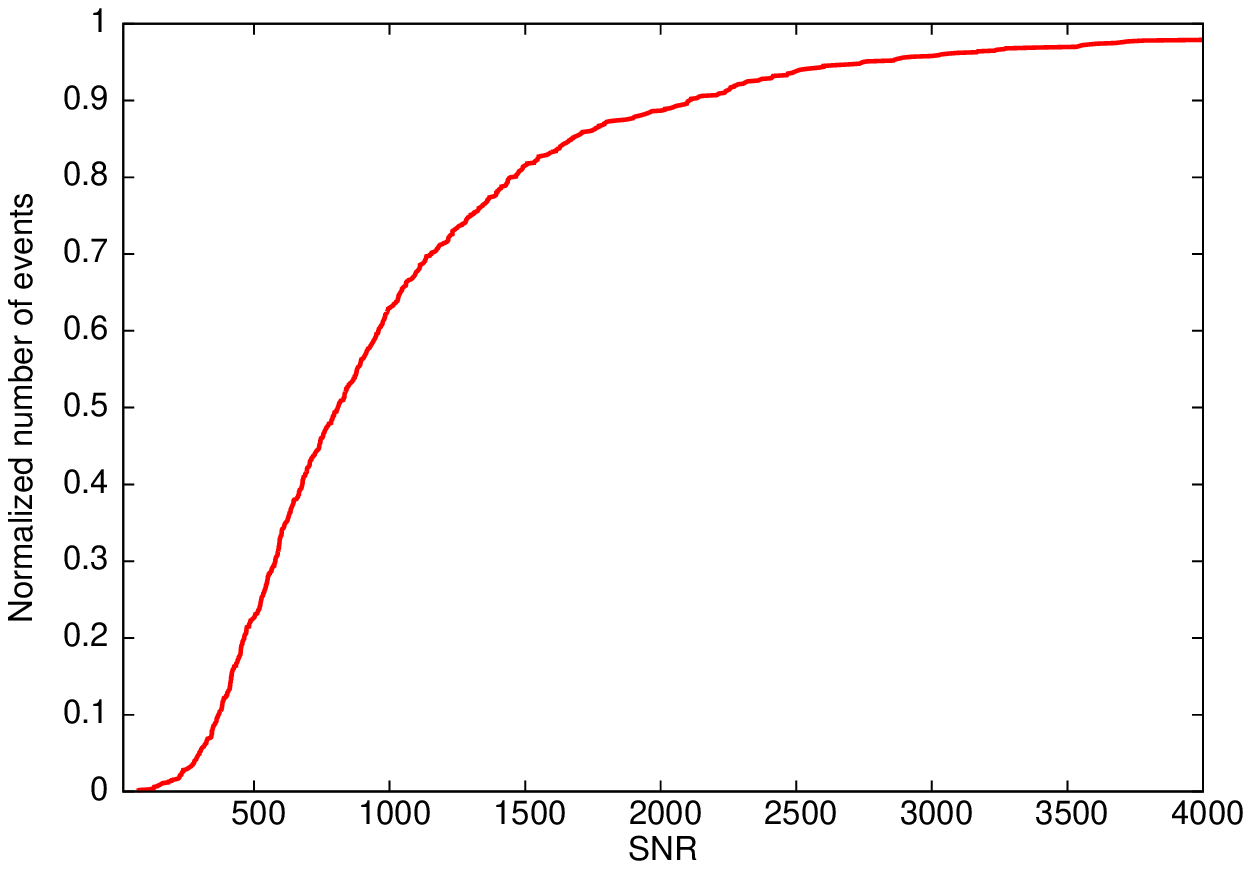}
}
\centerline{
\includegraphics[height=0.33\textwidth,  clip]{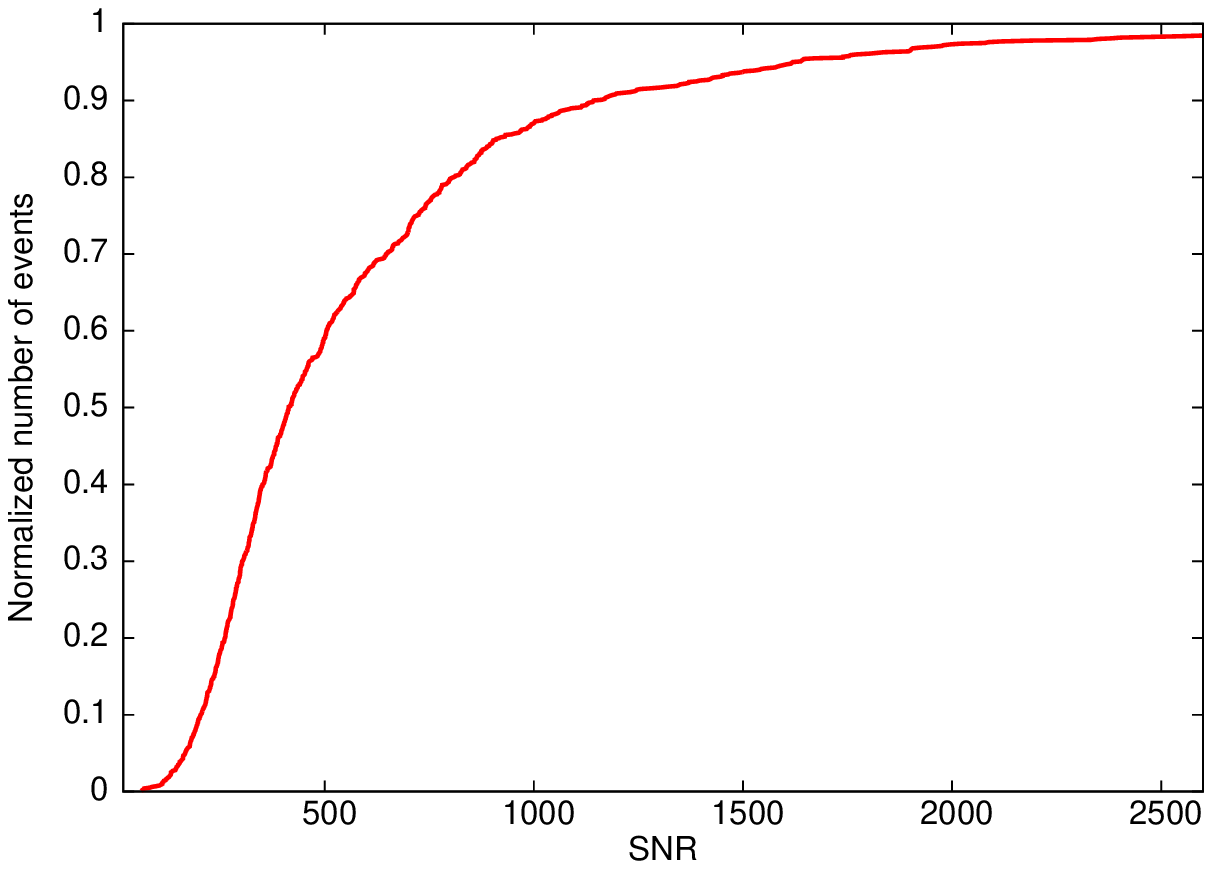}
\includegraphics[height=0.33\textwidth,  clip]{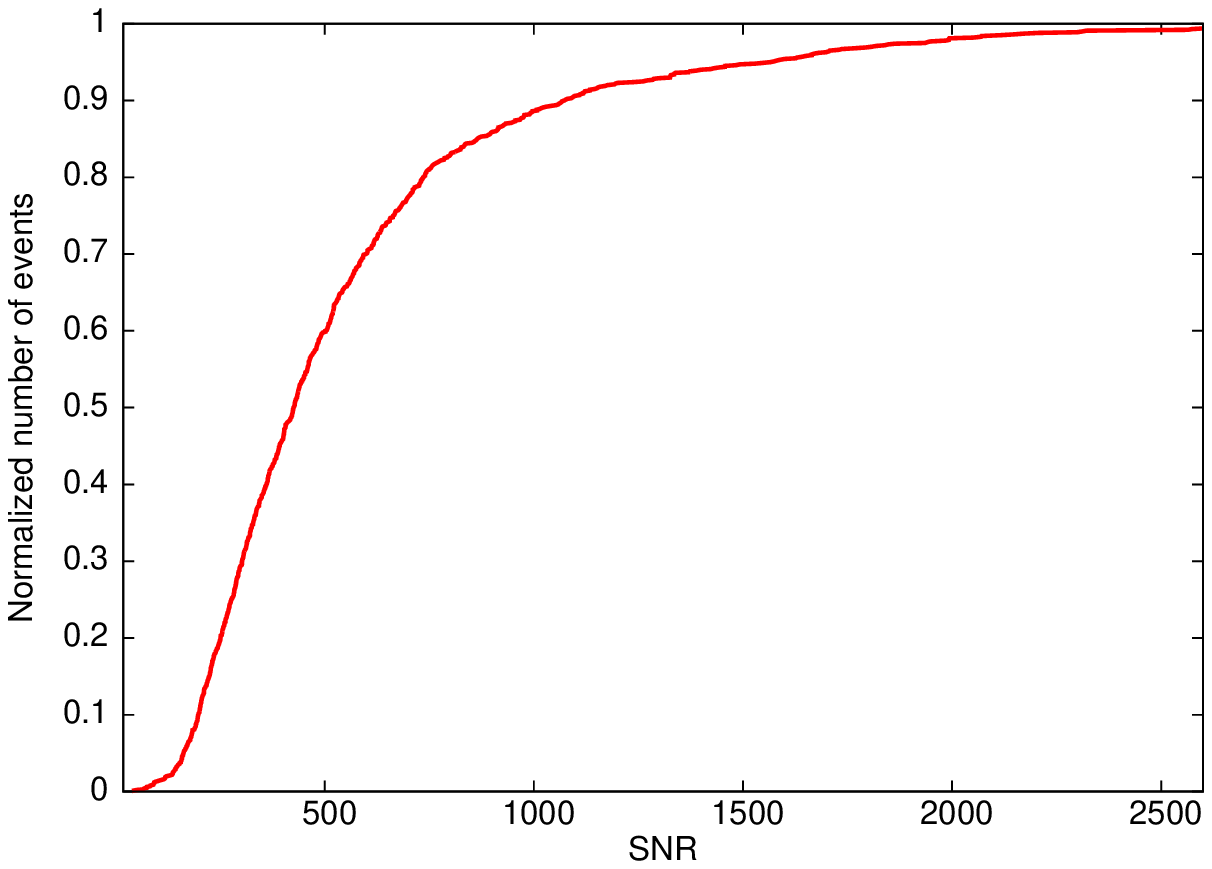}
}
\caption{Normalized cumulative distribution function for the signal--to--noise ratio of a cosmological population of binary systems with BHs of redshifted mass  $5\times10^3M_{\odot}+10^6M_{\odot}$. The panels show results for three different combinations of the spin magnitudes of the central and inspiralling BHs, \(q,\chi\), namely: top panel \(q=0.9, \chi=0.1\); bottom-left panel \(q=0.1,\chi=0.9\); bottom-right panel \(q=0.1, \chi=0.1\).}
\label{snrs}
\end{figure*}

In the Monte Carlo simulation, the intrinsic parameters of the source were fixed, with redshifted masses equal to the values quoted earlier, e.g., $10M_\odot+10^6M_{\odot}$ etc. For each source, we computed the SNR accumulated in a one year observation prior to plunge. This was accomplished by choosing the initial radial coordinate, $p_0$, such that the inspiralling BH reached the last stable orbit after one year of inspiral.  Figure~\ref{smoothdis} presents the normalized cumulative distribution functions of the SNR computed in this way, for binary systems with central BHs of redshifted mass  $10^6M_{\odot}$, and spin parameter $q=0.9$. The inspiralling BHs have specific spin parameter \(\chi =0.9\), and redshifted masses $\mu=10M_{\odot}$, $\mu=10^2M_{\odot}$, $\mu=10^3M_{\odot}$, and $\mu=5\times10^3M_{\odot}$.  In Figure~\ref{snrs}, we present the normalized cumulative distribution function for  binary systems with redshifted masses $5\times10^3M_{\odot}+10^6M_{\odot}$, for three additional combinations of the spin magnitudes of the binary components. 

The way to interpret the normalised cumulative distributions of Figures~\ref{smoothdis}, \ref{snrs} is the following: the value of the cumulative distribution at a given \({\rm SNR_t}\), indicates the fraction of events with \({\rm SNR} < {\rm SNR_t}\).  Hence, for the least massive system, the cumulative distribution indicates that \(\sim 50\%\) of events in the Monte Carlo sample have  \({\rm SNR}\lesssim30\). This ``median'' SNR is a suitable reference SNR at which to quote parameter estimation accuracies for the binary systems considered in our studies. 

Based on Figures~\ref{smoothdis} and \ref{snrs}, we will therefore normalise our results to the following reference SNRs: for the inspiralling BHs with masses $\mu=10M_{\odot},\, 10^2M_{\odot},\, 10^3M_{\odot}, 5\times10^3M_{\odot}$, and spin magnitude \(\chi=0.9\), we will use a reference SNR of $30, 150, 400, 1000$, respectively. For the more massive system, $5\times10^3M_{\odot}+10^6M_{\odot}$, Figure~\ref{snrs} suggests that for the combination of spin magnitudes \(q=0.9, \chi=0.1\), we should use a reference SNR of $1000$; whereas for the combination \(q=0.1, \chi=0.9\), and \(q=0.1, \chi=0.1\), we should use SNR\(=500\). 

The results shown in Figures~\ref{smoothdis} and \ref{snrs} provide information that could be used to renormalise the results presented later in the paper, in Tables~\ref{tabFMErrBH}-\ref{tabFMErrMMBHs}, to a cosmological population of sources at $z \lesssim 1$.

\subsection{Parameter estimation results}
The parameter space we have considered is 11-dimensional. Five of these are intrinsic parameters, namely \(\ln m, \ln M, q, \chi, p_0\), while the other six are extrinsic or phase parameters. We summarize the physical meaning of the parameters in Table \ref{tableparams}. 

\begin{table}[thb]
\centerline{$\begin{array}{c|l}\hline\hline
\ln \mu  &\text {mass of inspiralling object}   \\
\ln M    &\text {mass of central SMBH}\\
q        &\text{magnitude of (specific) spin angular momentum of SMBH} \\
\chi     &\text{magnitude of (specific) spin angular momentum of inspiralling object} \\
p_0      &\text{Initial radius of inspiralling object's orbit} \\
\phi_0   &\text{Initial phase of  inspiralling object's orbit}      \\
\theta_S &\text{source sky colatitude in an ecliptic--based system }  \\
\phi_S   &\text{source sky azimuth in an ecliptic--based system}  \\
\theta_K  &\text{direction  of SMBH spin (colatitude)}  \\
\phi_K   &\text{direction of SMBH spin (azimuth)}  \\
\ln D    &\text{distance to  source}\\
\hline\hline
\end{array}$}
\caption{\protect\footnotesize
This table describes the meaning of the parameters used in our model. The angles ($\theta_S$,\,$\phi_S$) and ($\theta_K$,\,$\phi_K$) are defined in a fixed ecliptic--based coordinate system.}
\label{tableparams}
\end{table}

In Tables \ref{tabFMErrBH}, \ref{tabFMErrmBH}, \ref{tabFMErrMBH}, \ref{tabFMErrMMBH} and \ref{tabFMErrMMBHs}, we summarise the results of Monte Carlo simulations of the parameter estimation errors, over possible values for the extrinsic parameters of the source, but with fixed intrinsic parameters. Each Table present results for a different binary system. All of the systems have a central black hole with mass $M=10^6M_{\odot}$ and spin parameter $q=0.9$, and an inspiraling BH with spin parameter \(\chi =0.9\), but we make four different choices for the mass of the inspiralling BH, $\mu=10M_{\odot},\, 10^2M_{\odot},\, 10^3M_{\odot},\, 5\times10^3M_{\odot}$. For completeness in the analysis, and to explore the trend with the spin of the small/big body, we consider three additional cases for the most massive inspiraling BH, in which we vary the spin parameters of the components, choosing \(q=0.9, \chi=0.1\); \(q=0.1, \chi=0.9\); and \(q=\chi=0.1\). These results are shown in Tables~\ref{tabFMErrMMBH} and \ref{tabFMErrMMBHs}.

\begin{table}[thb]
\begin{tabular}{|c|c|c|c|c|c|c|c|c|c|c|c|c|}
\hline\multicolumn{2}{|c|}{}&\multicolumn{11}{c|}{Distribution of \(\log_{10}(\Delta X)\) in error, \(\Delta X\), for parameter \(X=\)}\\\cline{3-13}
\multicolumn{2}{|c|}{Model}&$\ln(m)$&$\ln(M)$&$q$&$\chi$&$p_0$&$\phi_0$&$\theta_S$&$\phi_S$&$\theta_K$&$\phi_K$&$\ln(D)$\\\hline
&Mean             &-3.99&-3.58&-4.13&1.68 &-3.31&-0.99&-1.54&-1.58&-1.14&-1.07&-1.09\\\cline{2-13}
$q=0.9$&St. Dev.  &0.110&0.122&0.114&0.098&0.124&0.453&0.133&0.199&0.433&0.456&0.311\\\cline{2-13}
&L. Qt.           &-4.08&-3.61&-4.34&1.63 &-3.35&-1.35&-1.63&-1.71&-1.49&-1.40&-1.37\\\cline{2-13}
\(\chi=0.9\)&Med. &-3.99&-3.55&-4.08&1.68 &-3.29&-1.16&-1.53&-1.62&-1.24&-1.08&-1.16\\\cline{2-13}
&U. Qt.           &-3.89&-3.51&-3.95&1.74 &-3.23&-0.85&-1.49&-1.51&-0.85&-0.79&-0.88\\\hline
\end{tabular}
\caption{Summary of results of the Monte Carlo simulation of Fisher Matrix errors for spinning BH systems with specific spin parameters \(q=\chi=0.9\), and masses $\mu=10M_{\odot}$, \(M=10^6M_{\odot}\). We show the mean, standard deviation, median and quartiles of the distribution of the logarithm to base ten of the error in each parameter. Results are given for the kludge model with conservative corrections to 2PN order. The angles $\bar\phi_0$ and $\bar\alpha_0$, specifying LISA's position and orientation at $t=0$, are set to zero. Note that the results have been normalised to fixed SNR$=30$.}
\label{tabFMErrBH}
\end{table}

\begin{table}[thb]
\begin{tabular}{|c|c|c|c|c|c|c|c|c|c|c|c|c|}
\hline\multicolumn{2}{|c|}{}&\multicolumn{11}{c|}{Distribution of \(\log_{10}(\Delta X)\) in error, \(\Delta X\), for parameter \(X=\)}\\\cline{3-13}
\multicolumn{2}{|c|}{Model}&$\ln(m)$&$\ln(M)$&$q$&$\chi$&$p_0$&$\phi_0$&$\theta_S$&$\phi_S$&$\theta_K$&$\phi_K$&$\ln(D)$\\\hline
&Mean             &-3.78&-3.62&-4.98&0.19 &-2.76&-1.60&-2.04&-2.00&-1.81&-1.68&-1.78\\\cline{2-13}
$q=0.9$&St. Dev.  &0.074&0.075&0.112&0.068&0.075&0.330&0.206&0.246&0.361&0.396&0.270\\\cline{2-13}
&L. Qt.           &-3.84&-3.68&-5.06&0.14 &-2.82&-1.86&-2.20&-2.18&-2.08&-1.98&-2.02\\\cline{2-13}
\(\chi=0.9\)&Med. &-3.79&-3.63&-4.98&0.19 &-2.77&-1.66&-2.03&-2.01&-1.89&-1.74&-1.85\\\cline{2-13}
&U. Qt.           &-3.74&-3.57&-4.88&0.25 &-2.71&-1.43&-1.87&-1.88&-1.56&-1.40&-1.58\\\hline
\end{tabular}
\caption{ As Table~\ref{tabFMErrBH}, but for an inspiralling BH with mass $\mu=100M_{\odot}$. Results are quoted at a fixed SNR of 150.}
\label{tabFMErrmBH}
\end{table}

\begin{table}[thb]
\begin{tabular}{|c|c|c|c|c|c|c|c|c|c|c|c|c|}
\hline\multicolumn{2}{|c|}{}&\multicolumn{11}{c|}{Distribution of \(\log_{10}(\Delta X)\) in error, \(\Delta X\), for parameter \(X=\)}\\\cline{3-13}
\multicolumn{2}{|c|}{Model}&$\ln(m)$&$\ln(M)$&$q$&$\chi$&$p_0$&$\phi_0$&$\theta_S$&$\phi_S$&$\theta_K$&$\phi_K$&$\ln(D)$\\\hline
&Mean             &-3.33&-3.15&-4.62&-0.52&-2.03&-1.39&-1.88&-1.84&-1.67&-1.56&-1.73\\\cline{2-13}
$q=0.9$&St. Dev.   &0.096&0.097&0.091&0.084&0.097&0.407&0.344&0.363&0.444&0.470&0.321\\\cline{2-13}
&L. Qt.           &-3.41&-3.23&-4.70&-0.59&-2.11&-1.66&-2.15&-2.06&-1.95&-1.87&-1.94\\\cline{2-13}
\(\chi=0.9\)&Med. &-3.38&-3.20&-4.67&-0.55&-2.08&-1.50&-1.82&-1.79&-1.72&-1.62&-1.82\\\cline{2-13}
&U. Qt.           &-3.24&-3.06&-4.54&-0.45&-1.93&-1.26&-1.60&-1.68&-1.43&-1.30&-1.55\\\hline
\end{tabular}
\caption{ As Table~\ref{tabFMErrBH}, but for an inspiralling BH with mass $\mu=10^3M_{\odot}$. Results are quoted at a fixed SNR of 400. }
\label{tabFMErrMBH}
\end{table}

\begin{table}[thb]
\begin{tabular}{|c|c|c|c|c|c|c|c|c|c|c|c|c|}
\hline\multicolumn{2}{|c|}{}&\multicolumn{11}{c|}{Distribution of \(\log_{10}(\Delta X)\) in error, \(\Delta X\), for parameter \(X=\)}\\\cline{3-13}
\multicolumn{2}{|c|}{Model}&$\ln(m)$&$\ln(M)$&$q$&$\chi$&$p_0$&$\phi_0$&$\theta_S$&$\phi_S$&$\theta_K$&$\phi_K$&$\ln(D)$\\\hline
&Mean              &-3.12&-2.94&-4.38&-1.07&-1.60&-1.48&-2.06&-2.07&-1.85&-1.71&-1.90\\\cline{2-13}
$q=0.9$&St. Dev.   &0.089&0.089&0.083&0.081&0.089&0.417&0.379&0.391&0.434&0.468&0.326\\\cline{2-13}
&L. Qt.            &-3.14&-2.99&-4.40&-1.09&-1.65&-1.91&-2.39&-2.23&-2.13&-2.00&-2.10\\\cline{2-13}
\(\chi=0.9\)&Med.  &-3.13&-2.94&-4.38&-1.06&-1.64&-1.72&-2.07&-2.02&-1.89&-1.73&-1.95\\\cline{2-13}
&U. Qt.            &-3.03&-2.84&-4.37&-1.02&-1.54&-1.42&-1.87&-1.74&-1.62&-1.53&-1.71\\\hline

&Mean              &-2.62&-2.44&-3.92&-0.60&-1.19&-1.49&-2.09&-2.04&-1.75&-1.62&-1.81\\\cline{2-13}
$q=0.9$&St. Dev.   &0.181&0.185&0.176&0.172&0.205&0.359&0.440&0.365&0.398&0.444&0.411\\\cline{2-13}
&L. Qt.            &-2.77&-2.59&-4.07&-0.70&-1.31&-1.96&-2.44&-2.32&-2.12&-2.09&-2.10\\\cline{2-13}
\(\chi=0.1\)&Med.  &-2.60&-2.41&-3.90&-0.61&-1.21&-1.66&-2.06&-2.03&-1.82&-1.77&-1.89\\\cline{2-13}
&U. Qt.            &-2.46&-2.29&-3.77&-0.44&-1.09&-1.36&-1.78&-1.69&-1.60&-1.41&-1.67\\\hline
\end{tabular}
\caption{ As Table~\ref{tabFMErrBH}, but for an inspiralling BH with mass $\mu=5\times10^3M_{\odot}$. Results are quoted at a fixed SNR of 1000.}
\label{tabFMErrMMBH}
\end{table}

\begin{table}[thb]
\begin{tabular}{|c|c|c|c|c|c|c|c|c|c|c|c|c|}
\hline\multicolumn{2}{|c|}{}&\multicolumn{11}{c|}{Distribution of \(\log_{10}(\Delta X)\) in error, \(\Delta X\), for parameter \(X=\)}\\\cline{3-13}
\multicolumn{2}{|c|}{Model}&$\ln(m)$&$\ln(M)$&$q$&$\chi$&$p_0$&$\phi_0$&$\theta_S$&$\phi_S$&$\theta_K$&$\phi_K$&$\ln(D)$\\\hline
&Mean              &-3.09&-2.92&-2.61&-0.15&-1.62&-1.58&-1.99&-1.89&-1.75&-1.63&-1.79\\\cline{2-13}
$q=0.1$&St. Dev.   &0.071&0.072&0.064&0.062&0.072&0.365&0.401&0.416&0.387&0.416&0.283\\\cline{2-13}
&L. Qt.            &-3.16&-2.99&-2.66&-0.20&-1.68&-1.82&-2.25&-2.15&-2.00&-1.92&-1.98\\\cline{2-13}
\(\chi=0.9\)&Med.  &-3.09&-2.93&-2.62&-0.16&-1.62&-1.66&-1.87&-1.85&-1.77&-1.64&-1.83\\\cline{2-13}
&U. Qt.            &-3.01&-2.85&-2.56&-0.11&-1.56&-1.41&-1.67&-1.64&-1.54&-1.38&-1.62\\\hline

&Mean              &-3.09&-2.92&-2.61&-0.15&-1.62&-1.68&-1.99&-1.90&-1.85&-1.71&-1.86\\\cline{2-13}
$q=0.1$&St. Dev.   &0.069&0.070&0.064&0.060&0.070&0.282&0.397&0.444&0.348&0.385&0.254\\\cline{2-13}
&L. Qt.            &-3.15&-2.98&-2.66&-0.20&-1.68&-1.86&-1.21&-2.12&-2.07&-1.96&-2.01\\\cline{2-13}
\(\chi=0.1\)&Med.  &-3.09&-2.92&-2.61&-0.16&-1.62&-1.68&-1.89&-1.83&-1.83&-1.71&-1.89\\\cline{2-13}
&U. Qt.            &-3.03&-2.86&-2.56&-0.11&-1.56&-1.54&-1.68&-1.59&-1.61&-1.48&-1.72\\\hline
\end{tabular}
\caption{ As Table~\ref{tabFMErrMMBH}, but for a slowly rotating central BH with spin parameter \(q=0.1\). Results are quoted at a fixed SNR of 500.}
\label{tabFMErrMMBHs}
\end{table}

\begin{figure*}[ht]
\centerline{
\includegraphics[height=0.38\textwidth,angle=0,  clip]{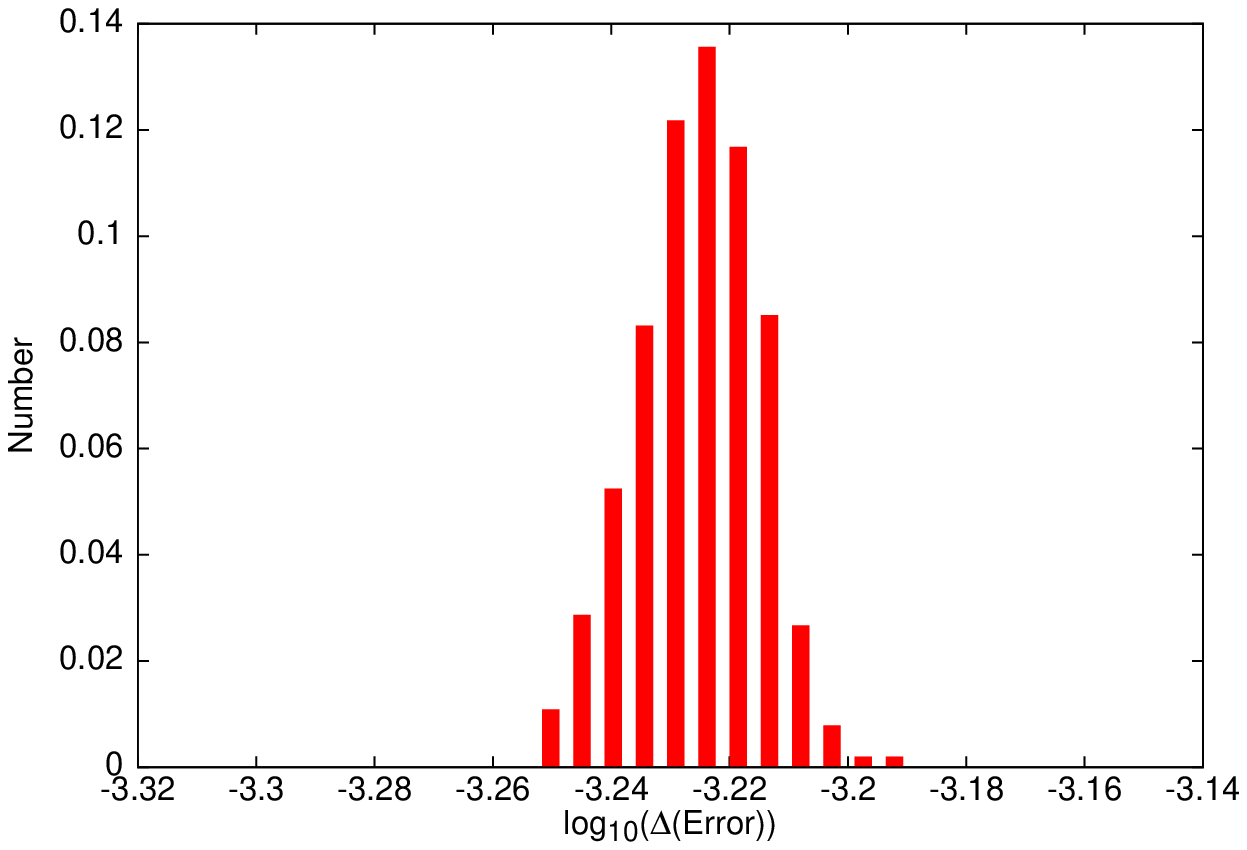}
\includegraphics[height=0.38\textwidth, angle=0,clip]{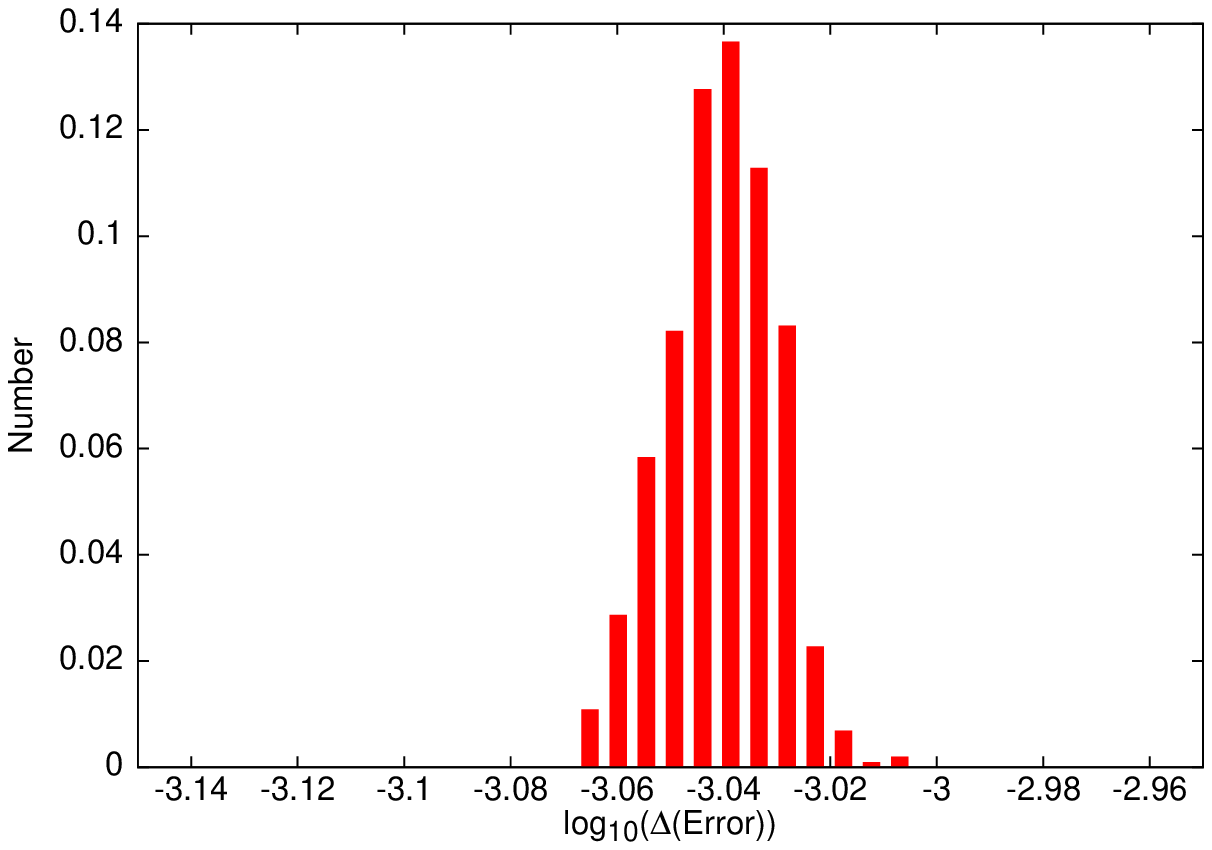}
}
\centerline{
\includegraphics[height=0.38\textwidth, angle=0, clip]{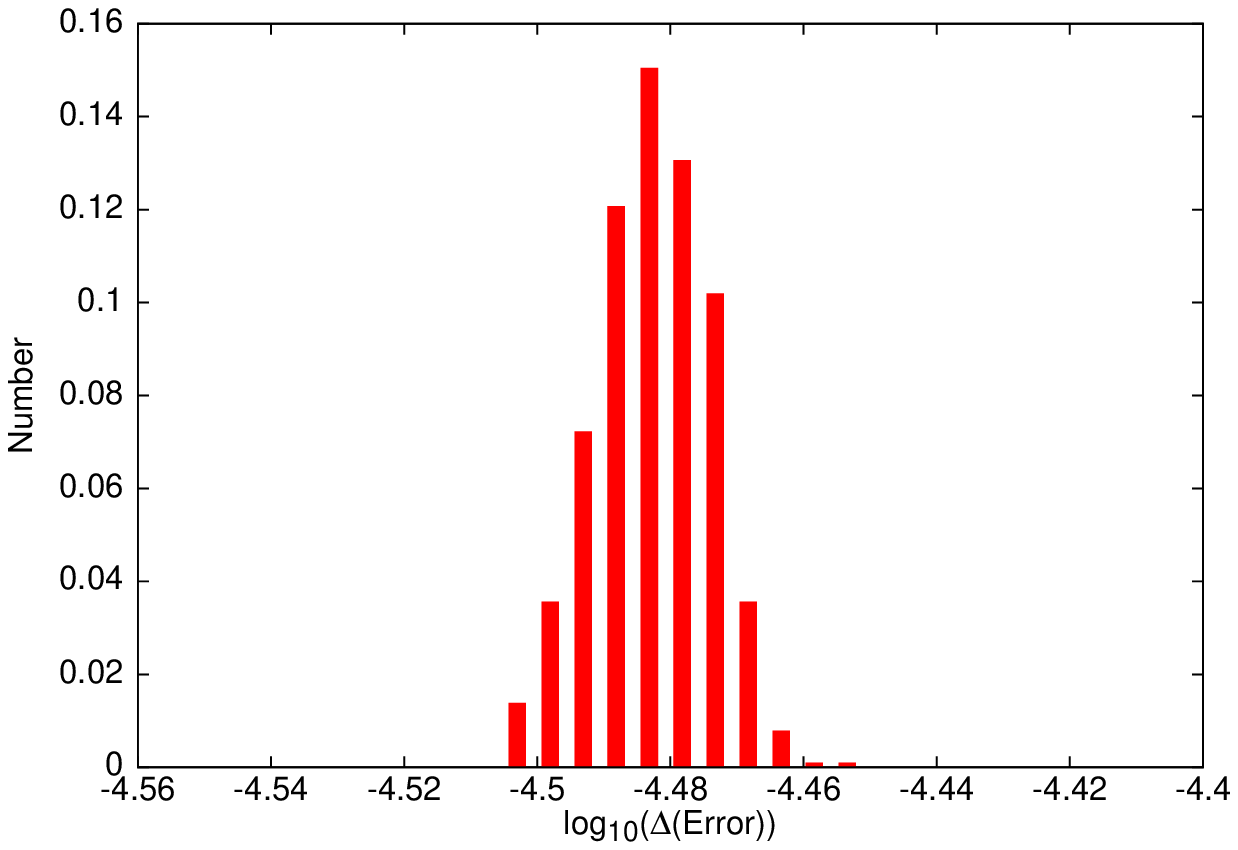}
\includegraphics[height=0.38\textwidth,angle=0, clip]{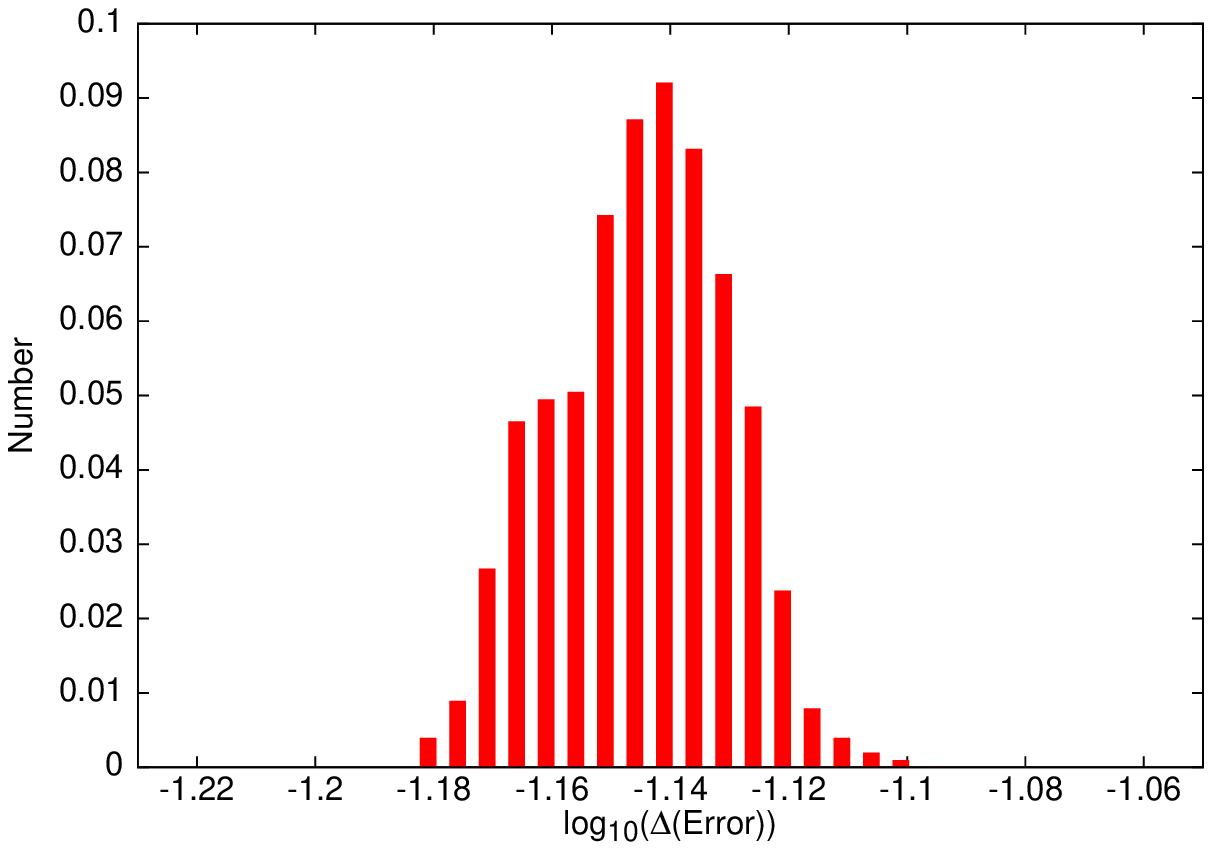}
}
\caption{Distribution of parameter measurement error estimates, \(\Delta\),  computed in the Monte Carlo simulations. We show results for the system with $\mu=5\times10^3M_{\odot}$, $M=10^6M_{\odot}$, $q=0.9$, \(\chi =0.9\). The panels show, from left to right, the error distributions for, top row:  $\Delta(\log_{10}(\ln \mu))$, $\Delta(\log_{10}(\ln M))$; and bottom row: $\Delta(\log_{10}(q))$ and $\Delta(\log_{10}(\chi))$ respectively. Results are quoted at fixed SNR of \(1000\).}
\label{mcfmdis}
\end{figure*}

The results in Table~\ref{tabFMErrBH} can be compared to out previous analysis for EMRIs that ignored the spin of the smaller black hole~\cite{cons}. We see that i) including the spin of the small CO for EMRIs with mass ratios \(\eta \lesssim 10^{-5}\) will not significantly affect parameter determination or detection and, ii) GW observations will not be able to constrain at all the spin parameter of the inspiralling BH for systems with such a small mass ratio. These results are consistent with the arguments presented by Barack \& Cutler \cite{cutler} in the sense that including small body spin effects for EMRIs has a minor effect on the orbital evolution of the system. 

For systems with mass ratios \(\eta \sim 10^{-4}\), Table~\ref{tabFMErrmBH} indicates that it will not be possible to measure the spin of inspiralling BHs in this regime either. However, Table~\ref{tabFMErrMBH} indicates that it should be possible to measure the spin of the inspiralling BH for systems with mass ratios \(\eta \sim 10^{-3}\). At an SNR of $400$, GW observations should be able to determine the spin parameter of the inspiralling BH for systems with component masses \(10^{3}M_{\odot} + 10^6 M_{\odot}\) to an accuracy of \(\sim30\%\). 

For inspiralling BHs of mass \(\mu=5\times10^3 M_{\odot}\), the results are quite similar, but the higher expected SNR for such systems should allow us to measure the small object spin to higher precision. For these systems, Table~\ref{tabFMErrMMBH} and Figure~\ref{mcfmdis} indicate that, at a fixed SNR$= 1000$, LISA measurements will be able to determine the inspiraling BH mass, SMBH mass, SMBH spin parameter, and inspiraling BH spin parameter with fractional errors of \(\sim 10^{-3},\,10^{-3},\,10^{-4},\,\textrm{and}\, \sim10^{-1}\). We also expect to determine the location of the source in the sky and the SMBH spin orientation to \(\sim10^{-4}\) steradians. 

We conclude that the inclusion of small body spin effects could become relevant for binaries with mass ratios \(\eta\gtrsim 10^{-3}\). Having found this threshold, we can explore the trend with the spin of the small/big body in this case. Tables~\ref{tabFMErrMMBH} and \ref{tabFMErrMMBHs} present four different combinations of the spin parameters of the more massive binary's components. For \(q=\chi=0.9\), Table~\ref{tabFMErrMMBH} suggests that GW observations will be able to measure the spin of the inspiraling body to a precision slightly better than ten percent. For \(q=0.9,\,\chi=0.1\), this same Table shows that when the inspiraling body is slowly rotating,  the accuracy with which we can determine the intrinsic parameters of the system is a factor of \(\sim 3\) times worse. However, even in this latter case, the small body spin magnitude could be measured to a fractional error of \(25\%\) and the accuracy with which the other extrinsic parameters can be determined remains basically unchanged. 

Table~\ref{tabFMErrMMBHs} shows that when the central SMBH is slowly rotating i) the accuracy with which the intrinsic parameters of the system can be determined is not very sensitive to the spin of the inspiraling body; ii) GW observations will not provide an accurate measurement of the spin of the inspiraling object. These results are to be expected --- the influence of spin couplings on the orbital evolution is what allows the spin of the inspiraling object to be measured from the gravitational wave emission. These spin couplings are enhanced when the two objects are more rapidly rotating, and when the inspiraling object is in the strong-field region close to the central massive object. As the innermost stable orbit is closer to the central SMBH when the latter is more rapidly rotating, this further increases the measurability of the spin coupling in that case. We conclude that the inspiraling object's spin will be most measurable when both bodies are rapidly rotating, which is supported by the results in Tables~\ref{tabFMErrMMBH} and \ref{tabFMErrMMBHs}. We see that GW observations in the low--frequency band could provide useful information not only on the spin distribution of SMBHs, but also of intermediate mass BHs that exist in the centres of galaxies, if binaries comprising two rapidly spinning objects are observed with sufficient SNR.

One final thing to note from Tables~\ref{tabFMErrBH}-\ref{tabFMErrMMBHs} is that the precision with which the other intrinsic parameters can be measured decreases as the mass of the inspiraling object increases. This is true even though the reference SNR used for each table is increasing, and would be even more pronounced if the results were renormalised to the same reference SNR for all systems. This trend is due to the fact that the more massive the binary system is, the more rapid the inspiral and the less time the inspiralling object spends close to the central SMBH. For the system \(\mu=10 M_{\odot}+ 10^6M_{\odot}\) with \(q=\chi=0.9\), the smaller object spends the last six months of inspiral within the region  \(p\in (\sim 6.01M \rightarrow \sim 2.32M)\). In contrast, for the system \(\mu=5\times 10^3 M_{\odot}+ 10^6M_{\odot}\) with \(q=\chi=0.9\), the object is at $p \approx 29.5M$ six months before plunge and hence spends much less time close to the innermost stable circular orbit (ISCO) of the central BH. Furthermore, the number of cycles completed over the last year of inspiral by the latter system is only  \(\sim10\%\) of the number of cycles completed by the former. 

\section{Model--induced parameter errors}
\label{s6}
The Fisher Matrix formalism we described in Section~\ref{s4} allowed us to estimate parameter estimation errors that would arise from noise in the detector. We shall now explore errors that would arise from using an approximate waveform model for data analysis. Even in the absence of noise, for a particular true signal, \({\bf s(t)}\), the kludge waveform model, \({\bf h(t)}\), that is the best match may have different parameters to the true waveform, which introduces another parameter error.  These are the ``model'' errors. Using the framework developed by Cutler and Vallisneri \cite{vallisneri}, and our own analysis on model errors for EMRIs in the context of spinless particles \cite{cons}, we will present results on the magnitude of the model errors that could arise in E/IMRIs in which the inspiral component has significant spin. 

As discussed by Cutler \& Vallisneri \cite{vallisneri}, we have no way to know what the ``true'' signal is. However, we can use their formalism  to estimate how important for parameter estimation the inclusion of various terms in the waveform model may be. In this section, we look specifically at what effect ignoring various ``conservative'' post-Newtonian terms will have on parameter determination. One feature that distinguishes ``model'' errors from noise--induced errors is that whereas the former are SNR independent, the latter scale as \(1/\textrm{SNR}\) (see \eqref{29}). Hence, it might be the case that for the systems with highest SNR, theoretical errors could dominate the total parameter--estimation error. In order to find out whether including conservative corrections is important for detection and parameter estimation, we will build waveform templates that include all, only part, or none of the conservative pieces we derived in Section~\ref{s2}. The kludge waveform template that includes all terms at 2PN order, see  Eq.~\eqref{omegacc}, will be taken as the ``true'' waveforms \(h_{\rm {\bf GR}}\). This model includes all the coefficients \(d_0, d_1...,\) etc.,  of the azimuthal frequency given in Eq.~\eqref{omegacc}. We then estimate the model errors by searching for \(h_{\rm {\bf GR}}\) using templates,  \(h_{\rm {\bf AP}}\), that include none or only part of these terms. When we refer to a template including conservative corrections at 1.5PN order, we will mean a template in which the angular frequency is given by Eq.~\eqref{omegacc} including the coefficients \(d_0, d_1, d_{1.5}, f_{1.5}\) (note that we found \(g_{1.5}=0\)). Similarly a template including conservative corrections at 0PN order, will mean a template in which all the coefficients except the leading $1$ inside the curly bracket of Eq.~\eqref{omegacc} are set to zero. We refer to this as testing the importance of including conservative corrections, as we will only alter the terms entering expression~\eqref{omegacc} for the frequency, and will not alter the radiative part of the waveform model described by the radial evolution in Eq.~\eqref{new_Ldot}--\eqref{6}. The reason for treating these two parts of the model differently is that the latter, radiative, part of the evolution can be computed straightforwardly from solution of the Teukolsky equation, while a calculation of the conservative part of the evolution requires knowledge of the local self-force, which is much more difficult to compute and is presently unknown.

At the end of this section, we will also assess the overall importance of including the spin of the small body in the model by comparing a waveform with $\chi = 0$, to one in which $\chi$ is kept as a free parameter. This will directly test whether using templates that do not include the small body spin will be adequate for accurate parameter estimation from GW observations. The rule of thumb in both of these analyses will be that if omitting part of the model gives rise to a parameter error that is comparable to or smaller than the noise-induced error, then we can safely ignore that part of the model when constructing search templates.

In order to provide an analysis as self--contained as possible, we shall now present a brief overview of the Cutler \& Vallisneri \cite{vallisneri} model error formalism that will be used for our subsequent analysis. Consider two manifolds embedded in the vector spaces of data streams. One of them is covered by the true waveforms \(\{h_{\rm GR}(\theta^i)\}\) and the other one by the approximate waveforms \(\{h_{\rm AP}(\theta^i)\}\). Given a signal \({\bf s} = {\bf h}_\mathrm{GR}(\hat \theta^i) + {\bf n}\), the best fit  \(\theta^i\) is  determined by the condition 

\begin{equation}
\label{36}
\Big( \partial_j \mathbf{h}_\mathrm{AP}(\theta^i) \Big|\mathbf{s} - \mathbf{h}_\mathrm{AP}(\theta^i) \Big) = 0.
\end{equation} 

\noindent Furthermore, at first order in the error \(\Delta\theta^i ({\bf 
n})\equiv \theta^i ({\bf n})- \hat \theta^i\), Eq. \eqref{36} takes the form

\begin{equation}
\Delta\theta^i =  \Big(\Gamma^{-1}(\tbf)\Big)^{ij} \, \Big( \partial_j \hAP(\tbf) \Big| \, \mathbf{n} \Big)
 +  \Big(\Gamma^{-1}(\tbf)\Big)^{ij} \, \Big( \partial_j \hAP(\tbf) \Big| \, \hGR(\ttr) - \hAP(\ttr) \Big) \, ,
\label{37}
\end{equation}
\noindent where the Fisher matrix is evaluated using the approximate waveforms \(\Gamma_{ij}(\tbf) \equiv ( \partial_i \hAP(\tbf)|\partial_j \hAP(\tbf))\). 

Relation \eqref{37} indicates that, at leading order, \(\Delta \theta^i\) is the sum of two contributions. The first one is due to noise in the detector, \(\Delta_n \theta^i\), whereas  the second one, \(\Delta_\mathrm{th}\theta^i\), is the contribution due to the inaccurate waveform model. These are given, respectively, by 
\begin{eqnarray}
\Delta_n\theta^i  & =&   \Big(\Gamma^{-1}(\tbf)\Big)^{ij} \Big( \partial_j \hAP(\tbf) \Big| \mathbf{n} \Big),  \qquad
\Delta_\mathrm{th}\theta^i =  \Big(\Gamma^{-1}(\tbf)\Big)^{ij} \Big( \partial_j \hAP(\tbf) \Big| \hGR(\ttr) - \hAP(\ttr)\Big). \label{38.2}
\end{eqnarray}
\noindent If we knew both \( \ttr\) and the noise realization \({\bf n}\), 
then these equations would allow us to determine \(\tbf\). However, experimentally we are only able to determine the  \(\hAP(\tbf)\) that is the best fit waveform for the data stream, \(\mathbf{s}\), and we are unsure about the error \(\Delta \theta \equiv \tbf - \ttr\). In addition, we do not know $\hat{\theta}$ in Eq.~\eqref{38.2}. At leading order, we can replace,  \(\hGR(\ttr) - \hAP(\ttr)\) by \(\hGR(\tbf) - \hAP(\tbf)\), obtaining
\begin{equation}
\label{39}
\Delta_\mathrm{th}\theta^i =  \Big(\Gamma^{-1}(\tbf)\Big)^{ij} \Big( \partial_j \hAP(\tbf) \Big| \hGR(\tbf) - \hAP(\tbf)\Big).
\end{equation}
\noindent This relation is both noise and SNR independent. This property, along with the fact that \(\Delta_\mathrm{th}\theta^i\) is not averaged out if the same event is measured by a large number of nearly identical detectors leads us to consider  \(\Delta_\mathrm{th}\theta^i\) as a systematic error.

\newcommand{\hla}{\mathbf{h}}
\newcommand{\tla}{\theta(\lambda)}
\newcommand{\ttl}{\theta_\mathrm{tr}(\lambda)}

Cutler and Vallisneri~\cite{vallisneri} found that the leading order approximation, Eq.~\eqref{39}, was not very good, unless the waveform is re--written in an amplitude-phase form
\begin{equation}
\label{40}
\tilde h^{\alpha}(f) =  A^{\alpha}(f) e^{i \Psi^{\alpha}(f)} \,.
\end{equation}

\noindent The amplitude \(A\) and phase \(\Psi\) are given by
\begin{eqnarray}
A_I &= & \frac{\sqrt{3} \, m}{2 \, D}\sqrt{A^{2}_{+} F_{I,+}^2 + A_{\times}^2 F_{I,\times}^2} \,, \qquad
\Psi_I =2 \phi + \psi_I, \\
A_{II} &= &\frac{\sqrt{3} \, m}{2 \, D} \sqrt{A^{2}_{+} F_{II,+}^2 + A_{\times}^2 F_{II,\times}^2} \,, \qquad
\Psi_{II} =2 \phi + \psi_{II},
\end{eqnarray}
\noindent where \(A_{+, \times}\) are given by \eqref{15} with \(A_{+}= A_{\times}= 2 p^2 \dot{\phi}^2\), \(F_{\{I, II\,;\,+, \times\}}\)  are given by \eqref{16} and \eqref{17}, and 
\begin{eqnarray}
\psi_I &=& \arctan\left(- \frac{A_{\times}F_{I,\times}}{A_{+}F_{I,+}}\right),\qquad
\psi_{II} = \arctan\left(- \frac{A_{\times}F_{II,\times}}{A_{+}F_{II,+}}\right).
\label{phases}
\end{eqnarray}
\noindent The first order approximation to this expression
\begin{equation}
\label{42}
\Delta_\mathrm{th}\theta^i \approx
\big(\Gamma^{-1}(\tbf) \big)^{ij}  \Big( \underbrace{\big[ \Delta {\bf A} + i {\bf A} \Delta \boldsymbol{\Psi} \big] e^{i \boldsymbol{\Psi}}}_{\text{at}\,\theta} \Big| \,
  \partial_j \hAP(\tbf) \Big),
\end{equation}

\noindent was found to give reliable results when compared to more accurately computed error estimates~\cite{vallisneri}, so we use this form again here. Equation~\eqref{42} behaves better than Eq.~\eqref{39} since the difference between two waveforms, \(\hAP(\tbf)  - \hAP(\ttr)\), is not very well approximated by the first term in its Taylor expansion. The differences in both the amplitude and phase of the waveform are individually well approximated by the linear terms in the Taylor series \cite{vallisneri}. In fact, \eqref{39} is reliable only as long as the phase difference between the two waveforms is much less than one radian, i.e. \( \Delta \theta^j \partial_j \boldsymbol{\Psi}_\mathrm{AP}(\tbf) \ll 1\), whereas for \eqref{42} we just require \(\Delta \theta^i \Delta \theta^j \partial_i \partial_j \boldsymbol{\Psi}_\mathrm{AP}(\tbf) \ll 1\). This condition is much less restrictive than the former one.

We can now use equation \eqref{42} to estimate the magnitude of the parameter errors that arise from inaccuracies in the template waveform.  In \cite{cons} we introduced a waveform model that included conservative self--force corrections at 2PN for Kerr circular--equatorial EMRIs. The conservative corrections used in that model were derived using the same method employed in Section~\ref{s2}, since accurate fully relativistic corrections are not currently available for objects moving in a Kerr background. There is, however, an ongoing program that aims at developing accurate and efficient numerical methods for computing such corrections. As mentioned in Section~\ref{s5},  first-order self--force results have been derived for a scalar charge moving on circular and eccentric equatorial orbits in the Kerr  spacetime~\cite{warleor,war}. The extension of this analysis to the gravitational case is now under investigation.  The computation of conservative corrections for the case we consider in this paper, namely, spinning BHs moving in the background of a Kerr BH, is beyond the scope of the current self--force program, which is focused on inspiraling objects that are non-spinning.

In the absence of fully perturbative calculations, we can use the kludge model to assess the importance of conservative corrections for detection and parameter estimation. This investigation will indicate whether it is necessary, for source detection and parameter estimation, to know these corrections accurately before we can properly analyse data from a space-based gravitational wave detector. This is an important question considering that these corrections are not yet known. 

We will compute  the model error that arises when we omit some or all of the ``conservative'' corrections in the waveform template and hence calculate the ratio \(\cal{R}\) of this model error to the error that will arise from noise in the detector. This ratio will indicate the importance of including the conservative corrections for parameter determination. If \(\cal{R}\) \(\lesssim 1\), then parameters estimated using a model that ignores the conservative corrections should still be reliable, but if \(\cal{R}\) \(>>1\) then it is clear that we must include the conservative corrections. The ratios of parameter errors to Fisher Matrix errors obtained from the Monte Carlo simulations over extrinsic parameters are summarized in Tables~\ref{tabFMRatBH}, \ref{tabFMRatmBH}, \ref{tabFMRatMBH}, \ref{tabFMRatMMBH}, \ref{tabFMRatMMBHslow}, and \ref{chivsnochi}.

We quote results for the same four test systems that we used previously, and consider two different comparisons, namely, we take the ``true'' waveform to be our kludge waveform with 2PN conservative corrections, and the template to be a kludge waveform with either no conservative corrections (``$0$PN'') or with conservative corrections to $1.5$PN order (``$1.5$PN''), and do pairwise comparisons. The exact meaning of $0$PN etc. was described in detail at the start of this section.

\begin{table}[thb]
\begin{tabular}{|c|c|c|c|c|c|c|c|c|c|c|c|c|}
\hline\multicolumn{2}{|c|}{}&\multicolumn{11}{c|}{\small{\(\log_{10}\) of the ratio \(\cal{R}\) of model to noise--induced error for parameter \(X=\)}}\\\cline{3-13}
\multicolumn{2}{|c|}{Model}&$\ln(m)$&$\ln(M)$&$q$&$\chi$&$p_0$&$\phi_0$&$\theta_S$&$\phi_S$&$\theta_K$&$\phi_K$&$\ln(D)$\\\hline
&Mean           &-0.50 &-0.26 &-0.31 &-0.35  &-0.26  &-0.28 &-0.30 &-0.37 &-0.12 &-0.09 &-0.16\\\cline{2-13}
&St. Dev.       &0.583 &0.441 &0.590 &0.622  &0.481  &0.778 &0.606 &0.623 &0.703 &0.718 &0.824\\\cline{2-13}
&L. Qt.         &-0.85 &-0.60 &-0.70 &-0.74  &-0.59  &-0.54 &-0.56 &-0.61 &-0.47 &-0.46 &-0.56\\\cline{2-13}
2PN vs 0PN&Med.   &-0.42 &-0.15 &-0.20 &-0.17  &-0.15  &-0.15 &-0.16 &-0.23 &0.04  & 0.03 &0.02\\\cline{2-13}
&U. Qt.         &-0.09 &0.17  &0.11  &0.15   &0.19   &0.16  &0.10  & 0.09 &0.38  &0.36  &0.43\\\hline

&Mean           &-0.70 &-0.49 &-0.47 &-0.52  &-0.48 &-0.42 &-0.53 &-0.62 &-0.32 &-0.27 &-0.42\\\cline{2-13}
&St. Dev.       &0.578 &0.564 &0.603 &0.626  &0.610 &0.720 &0.619 &0.617 &0.763 &0.740 &0.813\\\cline{2-13}
&L. Qt.         &-0.98 &-0.73 &-0.77 &-0.79  &-0.72 &-0.65 &-0.72 &-0.83 &-0.60 &-0.51 &-0.68\\\cline{2-13}
2PN vs 1.5PN&Med. &-0.59 &-0.35 &-0.37 &-0.38  &-0.32 &-0.28 &-0.35 &-0.44 &-0.19 &-0.09 &-0.17\\\cline{2-13}
&U. Qt.         &-0.31 &-0.02 &-0.01 &-0.05  & 0.00 &-0.01 &-0.13 &-0.15 &0.14  &0.20  &0.16\\\hline
\end{tabular}
\caption{Summary of Monte Carlo simulation results for the ratio of model errors to noise-induced errors, computed using the Fisher Matrix, for spinning BH systems with $\mu=10M_{\odot}$. We show the mean, standard deviation, median and quartiles of the distribution of the logarithm to base ten of the ratio for each parameter. Results are given for various comparisons, as indicated and described in the text. A comparison ``A vs B'' uses model A as the true waveform, and model B as the search template. Note that the noise--induced errors are quoted at a fixed SNR\(=30\).}
\label{tabFMRatBH}
\end{table}

\begin{table}[thb]
\begin{tabular}{|c|c|c|c|c|c|c|c|c|c|c|c|c|}
\hline\multicolumn{2}{|c|}{}&\multicolumn{11}{c|}{\small{\(\log_{10}\) of the ratio \(\cal{R}\) of model to noise--induced error for parameter \(X=\)}}\\\cline{3-13}
\multicolumn{2}{|c|}{Model}&$\ln(m)$&$\ln(M)$&$q$&$\chi$&$p_0$&$\phi_0$&$\theta_S$&$\phi_S$&$\theta_K$&$\phi_K$&$\ln(D)$\\\hline
&Mean           &0.33  &0.37  &0.65  &0.47  &0.32  &0.62   &0.48  &0.53  &0.65  &0.65  &0.61\\\cline{2-13}
&St. Dev.       &0.645 &0.561 &0.555 &0.638 &0.564 &0.738  &0.654 &0.657 &0.710 &0.707 &0.659\\\cline{2-13}
&L. Qt.         &0.07  &0.05  &0.37  &0.16  &0.05  &0.31   &0.18  &0.29  &0.23  &0.29  &0.08\\\cline{2-13}
2PN vs 0PN&Med. &0.49  &0.50  &0.71  &0.67  &0.48  &0.79   &0.63  &0.67  &0.83  &0.82  &0.79\\\cline{2-13}
&U. Qt.         &0.74  &0.74  &1.06  &0.92  &0.74  &1.08   &1.04  &0.98  &1.25  &1.21  &1.27\\\hline

&Mean            &0.16   &0.14  &0.60  &0.34  &0.09   &0.46  &0.42 &0.42   &0.61  &0.64  &0.56\\\cline{2-13}
&St. Dev.        &0.521  &0.564 &0.536 &0.667 &0.635  &0.743 &0.703 &0.692 &0.688 &0.662 &0.776\\\cline{2-13}
&L. Qt.          &-0.11  &-0.17 &0.24  &0.02  &-0.18  &0.09  &0.06 &0.03   &0.19  &0.23  &0.15\\\cline{2-13}
2PN vs 1.5PN&Med.&0.30   &0.30  &0.69  &0.48  &0.27   &0.61  &0.51 &0.55   &0.81  &0.79  &0.78\\\cline{2-13}
&U. Qt.          &0.58   &0.58  &1.01  &0.78  &0.60   &1.02  &0.84 &0.97   &1.12  &1.11  &1.17\\\hline
\end{tabular}
\caption{ As Table~\ref{tabFMRatBH}, but for a spinning BH with mass $\mu=100M_{\odot}$. Noise--induced errors are quoted at a fixed SNR\(=150\).}
\label{tabFMRatmBH}
\end{table}

\begin{table}[thb]
\begin{tabular}{|c|c|c|c|c|c|c|c|c|c|c|c|c|}
\hline\multicolumn{2}{|c|}{}&\multicolumn{11}{c|}{\small{\(\log_{10}\) of the ratio \(\cal{R}\) of model to noise--induced error for parameter \(X=\)}}\\\cline{3-13}
\multicolumn{2}{|c|}{Model}&$\ln(m)$&$\ln(M)$&$q$&$\chi$&$p_0$&$\phi_0$&$\theta_S$&$\phi_S$&$\theta_K$&$\phi_K$&$\ln(D)$\\\hline
&Mean              &1.28  &1.30  &1.50  &1.38  &1.30   &1.81  &1.63  &1.68  &1.90  &1.91  &1.90\\\cline{2-13}
&St. Dev.          &0.599 &0.651 &0.494 &0.651 &0.563  &0.715 &0.725 &0.643 &0.664 &0.673 &0.725\\\cline{2-13}
&L. Qt.            &0.89  &0.92  &1.22  &0.98  &0.93   &1.24  &1.15  &1.21  &1.26  &1.24  &1.21\\\cline{2-13}
2PN vs 0PN&Med.    &1.30  &1.40  &1.47  &1.51  &1.40   &1.79  &1.69  &1.76  &1.96  &1.99  &1.98\\\cline{2-13}
&U. Qt.            &1.75  &1.75  &1.80  &1.88  &1.76   &2.52  &2.23  &2.20  &2.67  &2.69  &2.75\\\hline

&Mean              &0.24  &0.24  &0.58  &0.44  &0.25   &0.66  &0.77   &0.69  &0.80  &0.84  &0.81\\\cline{2-13}
&St. Dev.          &0.629 &0.614 &0.519 &0.611 &0.596  &0.759 &0.724  &0.708 &0.740 &0.691 &0.737\\\cline{2-13}
&L. Qt.            &-0.25 &-0.24 &0.35  &-0.04 &-0.27  &0.04  &0.05   &0.09  &-0.03 &0.03  &-0.01\\\cline{2-13}
2PN vs 1.5PN&Med.  &0.41  &0.40  &0.57  &0.53  &0.44   &0.70  &0.91   &0.73  &0.94  &0.95  &0.92\\\cline{2-13}
&U. Qt.            &0.79  &0.77  &0.87  &0.97  &0.79   &1.36  &1.61   &1.44  &1.65  &1.75  &1.74\\\hline
\end{tabular}
\caption{ As Table~\ref{tabFMRatBH}, but for a spinning BH with mass $\mu=10^3M_{\odot}$. Noise--induced errors are quoted at a fixed SNR\(=400\).}
\label{tabFMRatMBH}
\end{table}

\begin{table}[thb]
\begin{tabular}{|c|c|c|c|c|c|c|c|c|c|c|c|c|}
\hline\multicolumn{2}{|c|}{}&\multicolumn{11}{c|}{\small{\(\log_{10}\) of the ratio \(\cal{R}\) of model to noise--induced error for parameter \(X=\)}}\\\cline{3-13}
\multicolumn{2}{|c|}{Model}&$\ln(m)$&$\ln(M)$&$q$&$\chi$&$p_0$&$\phi_0$&$\theta_S$&$\phi_S$&$\theta_K$&$\phi_K$&$\ln(D)$\\\hline
&Mean              &1.72    &1.73& 1.82&  1.83&  1.73&   2.47&  2.24&  2.22&  2.54&  2.54&  2.56\\\cline{2-13}
2PN vs 0PN&St. Dev.&0.709   &0.686&0.597& 0.614& 0.677&  0.708& 0.679& 0.731& 0.752& 0.789& 0.826\\\cline{2-13}
&L. Qt.            &1.29    &1.25& 1.52&  1.35&  1.29&   1.92&  1.74&  1.80&  1.90&  1.90&  1.91\\\cline{2-13}
\(\chi=0.9\)&Med.  &1.81    &1.78& 1.85&  1.92&  1.77&   2.43&  2.24&  2.33&  2.56&  2.56&  2.54\\\cline{2-13}
&U. Qt.            &2.21    &2.21& 2.18&  2.32&  2.21&   3.04&  2.64&  2.81&  3.29&  3.34&  3.35\\\hline

&Mean              &0.38    &0.39  &0.76  &0.49  &0.39   &1.04  &0.93  &0.90  &1.04  &1.11  &1.10\\\cline{2-13}
2PN vs 1.5PN&St. Dev.&0.719 &0.604 &0.490 &0.585 &0.594  &0.720 &0.715 &0.637 &0.799 &0.735 &0.816\\\cline{2-13}
&L. Qt.            &-0.18   &-0.16 &0.56  &-0.11 &-0.18  &0.40  &0.38  &0.30  &0.39  &-0.02 &0.35\\\cline{2-13}
\(\chi=0.9\)&Med.  &0.47    &0.45  &0.82  &0.64  &0.51   &0.96  &1.00  &0.98  &1.08  &1.17  &1.11\\\cline{2-13}
&U. Qt.            &0.90    &0.87  &1.04  &1.04  &0.84   &1.77  &1.54  &1.45  &1.78  &1.93  &1.94\\\hline

&Mean              &0.50    &0.50  &0.77  &0.63  &0.50   &1.06  &0.97  &0.92  &1.11  &1.14  &1.10\\\cline{2-13}
2PN vs 1.5PN&St. Dev.&0.708 &0.622 &0.405 &0.612 &0.621  &0.731 &0.632 &0.680 &0.701 &0.703 &0.795\\\cline{2-13}
&L. Qt.            &0.02    &0.01  &0.58  &0.14  &0.03   &0.39  &0.38  &0.28  &0.34  &0.36  &0.30\\\cline{2-13}
\(\chi=0.1\)&Med.  &0.63    &0.63  &0.78  &0.77  &0.62   &1.00  &1.00  &0.97  &1.12  &1.16  &1.18\\\cline{2-13}
&U. Qt.            &0.98    &0.98  &1.02  &1.11  &0.97   &1.73  &1.68  &1.60  &2.00  &2.01  &1.99\\\hline

\end{tabular}
\caption{ As Table~\ref{tabFMRatBH}, but for a spinning BH with mass $\mu=5\times10^3M_{\odot}$. Noise--induced errors are quoted at a fixed SNR\(=1000\).}
\label{tabFMRatMMBH}
\end{table}

\begin{table}[thb]
\begin{tabular}{|c|c|c|c|c|c|c|c|c|c|c|c|c|}
\hline\multicolumn{2}{|c|}{}&\multicolumn{11}{c|}{\small{\(\log_{10}\) of the ratio \(\cal{R}\) of model to noise--induced error for parameter \(X=\)}}\\\cline{3-13}
\multicolumn{2}{|c|}{Model}&$\ln(m)$&$\ln(M)$&$q$&$\chi$&$p_0$&$\phi_0$&$\theta_S$&$\phi_S$&$\theta_K$&$\phi_K$&$\ln(D)$\\\hline
&Mean               &0.41    &0.40  &0.49  &0.49  &0.38   &0.79 &0.66  &0.62  &0.88  &0.91  &0.87\\\cline{2-13}
2PN vs 1.5PN&St. Dev. &0.597 &0.615 &0.517 &0.643 &0.607  &0.698 &0.739 &0.723 &0.776 &0.801 &0.799\\\cline{2-13}
&L. Qt.             &0.05    &0.05  &0.15  &0.13  &0.02   &0.22  &0.17  &0.09  &0.16  &0.21  &0.12\\\cline{2-13}
\(\chi=0.9\)&Med.   &0.46    &0.46  &0.60  &0.58  &0.45   &0.81  &0.73  &0.69  &0.93  &0.92  &0.99\\\cline{2-13}
&U. Qt.             &0.78    &0.75  &0.90  &0.87  &0.75   &1.47  &1.23  &1.19  &1.81  &1.75  &1.74\\\hline
&Mean               &0.43    &0.41&  0.51&  0.50&  0.41&  0.78&  0.66&  0.64&  0.90&  0.93&  0.92\\\cline{2-13}
2PN vs 1.5PN&St. Dev. &0.624 &0.619& 0.632& 0.681& 0.609& 0.702& 0.757& 0.695& 0.758& 0.749& 0.784\\\cline{2-13}
&L. Qt.             &0.06    &0.03&  0.17&  0.17&  0.03&  0.18&  0.20&  0.15&  0.18&  0.22&  0.19\\\cline{2-13}
\(\chi=0.1\)&Med.   &0.51    &0.48&  0.63&  0.64&  0.46&  0.79&  0.74&  0.74&  0.94&  1.00&  0.96\\\cline{2-13}
&U. Qt.             &0.81    &0.83&  0.95&  0.93&  0.83&  1.49&  1.28&  1.24&  1.79&  1.83&  1.85\\\hline
\end{tabular}
\caption{ As Table~\ref{tabFMRatMMBH}, but for a central SMBH with spin parameter \(q=0.1\). Noise--induced errors are quoted at a fixed SNR\(=500\).}
\label{tabFMRatMMBHslow}
\end{table}

\begin{table}[thb]
\begin{tabular}{|c|c|c|c|c|c|c|c|c|c|c|c|c|}
\hline\multicolumn{2}{|c|}{}&\multicolumn{11}{c|}{\small{\(\log_{10}\) of the ratio \(\cal{R}\) of model to noise--induced error for parameter \(X=\)}}\\\cline{3-13}
\multicolumn{2}{|c|}{Model}&$\ln(m)$&$\ln(M)$&$q$&$\chi$&$p_0$&$\phi_0$&$\theta_S$&$\phi_S$&$\theta_K$&$\phi_K$&$\ln(D)$\\\hline
&Mean                          &0.64  &1.12  &0.87  &6.21  &1.88  &3.41  &3.20  &3.21  &3.72  &3.98  &3.79 \\\cline{2-13}
\(\chi=0.9\)&St. Dev.          &0.519 &0.569 &0.610 &0.603 &0.629 &0.725 &0.617 &0.574 &0.697 &0.621 &0.649\\\cline{2-13}
vs&L. Qt.                      &0.37  &0.99  &0.63  &5.94  &1.70  &3.33  &3.00  &2.99  &3.31  &3.50  &3.34 \\\cline{2-13}
\(\chi=0\)&Med.                &0.73  &1.25  &1.02  &6.29  &1.98  &3.35  &3.29  &3.25  &3.81  &3.99  &3.85 \\\cline{2-13}
&U. Qt.                        &0.98  &1.43  &1.29  &6.58  &2.21  &3.55  &3.41  &3.44  &4.16  &4.33  &4.39 \\\hline
\end{tabular}
\caption{ Model errors that arise from omitting the spin of the inspiralling object for systems with component masses \([5\times 10^3+10^6] M_{\odot}\).}
\label{chivsnochi}
\end{table}

\noindent Tables~\ref{tabFMRatBH}--\ref{tabFMRatMMBHslow} indicate that, as expected, the ratio \(\cal{R}\) of model errors to noise--induced errors becomes smaller as the approximate waveform, \(\hAP\), becomes closer to the ``true'' waveform, \(\hGR\). The improvement going from $0$PN to $1.5$PN is largest for the more massive binaries.
 
From Table~\ref{tabFMRatBH}, we see that for the binary systems \(10M_{\odot}+10^6 M_{\odot}\), the vast majority of sources fulfill the condition that the model errors are smaller than the noise--induced errors, for the particular SNR, $30$, at which these results are quoted. We note also that the error ratio for the spin of the inspiralling object appears to be similar to the ratio for the other intrinsic parameters. However, this value is the ratio of two large numbers. As indicated in Section~\ref{s5}, the median of its noise-induced error is of the order of \(\sim10^{1.6}\), whereas its model error is of the order of \(\sim10^{1.2}\). Hence, although the ratio looks small, this is not an indication that this parameter can be accurately measured by EMRI observations.

Tables~\ref{tabFMRatmBH}--\ref{tabFMRatMMBHslow} indicate that, for the more massive BH binaries, model errors are likely to be larger than the statistical errors.  This result was already pointed out by Cutler \& Vallisneri \cite{vallisneri} in the context of the inspiral of non--spinning massive BH binaries. In their studies, they used a simple PN model with no spin corrections, for sources with a duration of 1 year, which were truncated at the ISCO of non--spinning BHs, i.e., \(p=6M\). We have now extended that initial calculation by i) building a more accurate waveform model using the equations of motion derived by Saijo et al. \cite{maeda} for spinning inspiralling objects; ii) augmenting this model with spin--orbit and spin--spin couplings taken from perturbative and PN calculations to amend the equations of motion; and iii) modelling the inspiral phase using the most accurate fluxes available which include perturbative corrections for the spin of the inspiralling body, spin--spin couplings and higher--order fits to Teukolsky calculations. Additionally, we have used a consistent prescription for the computation of the ISCO at which we truncate the waveforms. Small body spin effects produce fairly negligible changes in the value of ISCO for IMRIs, but we have included these in order to be consistent throughout our analysis.

There are two reasons why the model error to noise-error ratio appears to increase as the mass of the inspiralling body increases. Firstly, we are using a higher reference SNR to quote the noise errors for the more massive systems, since these are intrinsically louder. At higher SNR, the noise-induced error decreases while the model error is fixed. Nonetheless, if we renormalise the results in the tables to a fixed SNR, across all sources, we still see that  the error ratio, \(\cal{R}\), for the $2$PN to $0$PN comparison, tends to increase as the mass ratio \(\eta\) is increased. This is to be expected, since the corrections that we are omitting in the comparison are proportional to the mass-ratio and therefore should have a greater impact for higher mass-ratio systems.

However, Tables~\ref{tabFMRatmBH}--\ref{tabFMRatMMBHslow} also show that for the 2PN vs 1.5PN comparisons, the error ratio, \(\cal{R}\), increases, but remains small and if the results were renormalised to a fixed SNR, it would actually decrease slightly as the mass of the inspiraling object was increased. For the 2PN vs 1.5PN comparison with $\chi=0.9$ and $\mu=5\times10^3M_\odot$, the error ratio \({\cal{R}} \lesssim 4\) in all cases at a reference SNR of $1000$, which is already a manageably small value. At an SNR of $250$, the error ratio would be less than $1$.

In this same case, the value of \({\cal{R}}\) is a factor of \(\sim 20/30\) larger for the intrinsic/extrinsic parameters for the 2PN vs 0PN comparison relative to the 2PN vs 1.5PN comparison. Altogether, these results suggest that a model which included  conservative corrections up to 2PN order might be sufficient for parameter estimation, since the relative importance of the \(1.5{\rm PN}\rightarrow2{\rm PN}\) change is small, even for the most massive systems. More work is required to confirm this, by comparing the 2PN model to a higher order (3PN or 3.5PN) model, but the results presented here seem promising.

In Table~\ref{chivsnochi} we show the model errors that would arise if we did not include the spin of the inspiralling body in the waveform template, but it was included in the ``true'' waveform. We show results for the system with component masses  \([5\times 10^3+10^6] M_{\odot}\) since the spin has the biggest impact in that case. This exercise is important in order to understand how much the parameter estimation we could achieve from a GW observation would be degraded if we did not use spinning templates in our search. This table shows that not including small body spin corrections could significantly degrade the accuracy with which the CO mass, SMBH mass and SMBH spin may be determined, since the model errors associated with these parameters are a factor of \(\sim 5,\, 20,\, 10\), bigger than the noise-induced errors, respectively. We note that the large quoted error associated with the small body spin parameter \(\chi\), is merely an indication that we cannot determine that parameter using a waveform template that does not include it in the first place. These results suggest that the small body spin corrections in IMRIs should not be ignored in data analysis, since doing so may significantly degrade our determination of the system parameters.

In summary, this analysis shows that model errors could be a limiting factor in determining the system parameters, but only for IMRIs. Similarly, the small object spin should not be ignored in IMRI waveforms.  However, constructing a waveform template that included conservative
corrections up to 2PN order could be sufficient to reduce systematic errors
to an acceptable level. Such templates should be able to constrain a source to a sufficiently small region of the parameter space that a follow-up using more accurate and computationally expensive waveform templates would be possible, if such templates are available.   

\section{Conclusions}
\label{s7}
In this paper we have developed a kludge waveform model that includes small body spin corrections in order to explore the ability of a future low--frequency gravitational wave detector, such as LISA, to measure the spin of BHs inspiralling into much more massive Kerr BHs. This model uses the equations of motion derived by Saijo et al. \cite{maeda}, and includes first--order conservative self--force corrections to compute the evolution of the inspiralling object's orbital frequency. The trajectory of the inspiralling object is computed using prescriptions for the fluxes of energy and angular momentum \cite{improved}, augmented with perturbative results including spin--orbit and spin--spin couplings \cite{tanaka}.

This analysis has demonstrated that LISA observations will not be able to measure the spin of stellar mass COs inspiralling into SMBHs. This result is in accord with arguments presented elsewhere~\cite{cutler}. However, including small body spin effects could be relevant for signal detection and parameter estimation for binaries with mass ratio \(\eta \gtrsim 10^{-3}\). At a fixed SNR of $1000$, a LISA observation of a binary with masses \(5\times10^3M_{\odot}+10^6M_{\odot}\) whose components have specific spin parameter \(q=\chi=0.9\), will be able to determine the CO and SMBH masses, the SMBH spin magnitude and the inspiralling BH spin magnitude, \(\chi\), to within fractional errors of \(\sim 10^{-3},\,10^{-3},\,10^{-4}\), \(\sim 10\%\), respectively. We also expect to determine the location of the source in the sky and the SMBH spin orientation to within \(\sim 10^{-4}\) steradians.  Small body spin effects will be measurable when both components of the binary are rapidly rotating, but these effects will not be measurable or have a significant impact on parameter estimation when the central SMBH is slowly rotating. 

We have also studied in detail the importance, for parameter estimation, of including corrections that arise from the ``conservative'' part of the self--force, using the formalism developed by Cutler \& Vallisneri \cite{vallisneri}. We have found that, for a source with component masses \(5\times10^3M_{\odot}+10^6M_{\odot}\), and spin magnitudes \(q=\chi=0.9\), the relative error when using a \(1.5{\rm PN}\) template to detect a \(2{\rm PN}\) signal is small at fixed SNR, for all the systems we considered. Indeed, for the 2PN vs 1.5PN comparison, the ratio \({\cal{R}} \) of the model errors to the noise--induced errors is of order \({\cal{R}} \lesssim 4\) for the intrinsic parameters, even for the ``worst case'' of a \(5\times10^3M_{\odot}+10^6M_{\odot}\) binary detected with SNR of $1000$. In contrast, when comparing 0PN to  2PN templates, this error ratio \({\cal{R}} \) increases by a factor of \(\sim 20/30\) for intrinsic/extrinsic parameters. This suggests that including these corrections up to 2PN order might already be sufficient to reduce these systematic errors to an acceptable level. We also investigated the importance of including the spin of the small body in the waveform model. For systems with the most massive inspiraling objects, the model error that arises from omitting the spin could be one to two orders of magnitude larger than the noise-induced error, suggesting that including the spin in waveform templates will be important for such systems.

These results extend an analysis initially carried out by Cutler \& Vallisneri \cite{vallisneri}. Our studies include various ingredients that were not considered in there analysis. They used a purely post-Newtonian waveform, and compared a consistent $3$PN approximate waveform to a ``true'' $3.5$PN waveform. Moreover, their analysis included spin-orbit coupling as a model parameter, but the ``true'' waveforms were taken to be non-spinning in all cases, and they mostly focussed on near equal-mass systems. In our work, by contrast, we have focussed on more extreme-mass-ratio systems, we have included spin for the ``true'' waveforms, and our waveform model is based on accurate equations of motion for circular-equatorial spinning BH binary systems, which means it can be more readily applied in the strong-field. Finally, we have only considered the omission of ``conservative'' corrections when building our approximate templates, as these these are much more difficult to compute than the radiative corrections in the extreme-mass-ratio regime.

The \(5\times10^3M_{\odot}+10^6M_{\odot}\) system described here is an intermediate-mass-ratio inspiral. Another type of IMRI, in which a stellar-mass object ($\sim 1M_\odot$) falls into an IMBH ($\sim 1000M_\odot$), could be detected by future ground-based detectors, such as the Einstein Telescope~\cite{Freise:2009}. In~\cite{firstpaper,thirdpaper}, we assessed the capability of the Einstein Telescope to determine the parameters of such systems, when the inspiralling BH was non-spinning. The small-body-spin corrections in the model described here could be used to augment the IMRI model used in~\cite{firstpaper} to assess the importance and measurability of the small body spin effects in such systems. This should be studied in the future.

\section*{Acknowledgments}
EH is funded by CONACyT. JG's work is supported by the Royal Society. The Monte Carlo simulations described in this paper were performed using the Darwin Supercomputer of the University of Cambridge High Performance Computing Service (http://www.hpc.cam.ac.uk/), provided by Dell Inc. using Strategic Research Infrastructure Funding from the Higher Education Funding Council for England.

\bibliography{references}

\end{document}